\documentclass[useAMS,usenatbib,usegraphicx]{mn2e}
\bibpunct[, ]{}{}{;}{a}{,}{,}
\usepackage{graphicx}
\usepackage[dvipsnames]{color} 
\graphicspath{{plots/}}

\newcommand{\be}{\begin{equation}}
\newcommand{\ee}{\end{equation}}

\newcommand{\msun}{\mbox{M$_{\sun}$}}

\newcommand{\bl}[1]{\mbox{\boldmath$ #1 $}}
\def\la{\mathrel{\hbox{\rlap{\hbox{\lower4pt\hbox{$\sim$}}}\hbox{$<$}}}}
\def\ga{\mathrel{\hbox{\rlap{\hbox{\lower4pt\hbox{$\sim$}}}\hbox{$>$}}}}

\begin{document}

\title[Stellar velocity ellipsoids]{Shape and
orientation of stellar velocity ellipsoids in spiral galaxies}
\author[E. I. Vorobyov and Ch. Theis]{E. I. Vorobyov$^{1}$\thanks{E-mail:
vorobyov@astro.uwo.ca (EIV); theis@astro.univie.ac.at (ChT)} and Ch.
Theis$^{2}$ \\
$^{1}$Department of Physics and Astronomy, University of Western Ontario, 
London, Ontario, N6A 3K7, Canada and \\
Institute of Physics at South Federal University, Stachki 194, Rostov-on-Don, Russia
 \\
$^{2}$Institut f\"ur Astronomie, Universit\"at Wien,
              T\"urkenschanzstr. 17, 1180 Wien, Austria.}

\date{}

\maketitle

\label{firstpage}

\begin{abstract}
We present a numerical study of the properties of the stellar velocity distribution 
in stellar discs which have developed a saturated, two-armed spiral structure. 
We follow the growth of the spiral structure deeply into the non-linear regime
by solving the Boltzmann moment equations up to second order. By adopting the thin-disc 
approximation, we restrict our study of the stellar velocity distribution to the plane of 
the stellar disc. We find that the outer (convex) edges of stellar spiral arms
are characterized by peculiar properties of the stellar velocity ellipsoids, 
which make them distinct from most other galactic regions.
In particular, the ratio $\sigma_1:\sigma_2$ of the smallest versus largest 
principal axes of the stellar velocity ellipsoid can become abnormally small 
(as compared to the rest of the disc) near the outer edges of spiral arms.
Moreover, the epicycle approximation fails to reproduce the ratio $\sigma_{\phi\phi}:\sigma_{rr}$
of the tangential versus radial velocity dispersions in these regions. 
These peculiar properties of the stellar velocity distribution 
are caused by large-scale non-circular motions of stars,
which in turn are triggered by the non-axisymmetric gravitational field of 
stellar spiral arms.

The magnitude of the vertex deviation appears to correlate globally 
with the amplitude of the spiral stellar density perturbations. However, locally 
there is no simple correlation between the vertex deviation and the density perturbations.
In particular, the local vertex deviation does not correlate with the local gravitational 
potential and shows only a weak correlation with the local gravitational potential gradient.
We find that the local vertex deviation correlates best with the spatial gradients of mean
stellar velocities, in particular with the radial gradient of the mean radial velocity.
\end{abstract}


\section{Introduction}

Already since the end of the nineteenth century studies of stellar
kinematics in the solar neighbourhood revealed that the velocity
distribution of stars in galactic discs is non-isotropic (Kobold
\cite{kobold90}). An early interpretation by Kapteyn
(\cite{kapteyn05}) and Eddington (\cite{eddington06}) invoked a
superposition of (isotropic) stellar streams with different mean
velocities, by this creating an anisotropic velocity distribution.
However, an alternative interpretation by Schwarzschild
(\cite{schwarzschild07}) became the general framework for describing
the local velocity distribution of stars of equal age, though a
contamination by stellar streams is still considered to be important
for a proper analysis of velocity data. The Schwarzschild
distribution is based on a single but anisotropic ellipsoidal
distribution function. It is characterized by a Gaussian distribution
in all three directions $U$ (radial $r$), $V$ (tangential $\phi$) and
$W$ (vertical $z$) in velocity space, but with different velocity
dispersions $\sigma_{rr}$, $\sigma_{\phi\phi}$, and $\sigma_{zz}$. In
general, the velocity distribution is described by a velocity
dispersion tensor
\begin{equation}
  \sigma^2_{ij} \equiv \overline{(v_i - \bar{v_i})(v_j - \bar{v_j})},
  \label{mixed_vd}
\end{equation}
where $i$ and $j$ denote the different coordinate directions, i.e.\
radial ($r$), azimuthal ($\phi$) and vertical ($z$) direction
and $v_i$ gives the corresponding velocity. The bar denotes local 
averaging over velocity space{\footnote We emphasize that 
our definition of the velocity dispersion tensor is
different from the definition of e.g. Kuijken \& Tremaine (\cite{KT})
who do not use a square mark. Hence, our $\sigma^2_{ij}$ is equivalent to
Kuijken \& Tremaine's $\sigma_{ij}$.  }. 
The principal axes of this tensor form an imaginary surface that is called the
velocity ellipsoid. The principal axes of the velocity ellipsoid need not to be aligned with
the coordinate axes. In that case, the vertex deviation $l_{\rm v}$ (which is
defined as the angle between the direction in velocity space
pointing from the Sun to the Galactic centre and 
the direction of the $\sigma_{ij}$ major principal axis) does not
vanish.

The general trend for the velocity dispersions is that the radial
velocity dispersions are the largest and the vertical velocity
dispersions are the smallest among stars of same age and stellar type.
This trend holds when dispersions are averaged locally, i.e.\ in the
solar neighbourhood (Dehnen \& Binney \cite{Dehnen}), or globally over
the entire galactic discs. For instance, the ratio
$\sigma_{rr}:\sigma_{\phi\phi}:\sigma_{zz}$ of the radial, tangential,
and vertical stellar velocity dispersions in the disc of NGC~488 is
approximately $1:0.8:0.7$ (Gerssen et al.\ \cite{Gerssen}).

In the case of stationary, axisymmetric systems and appropriate
distribution functions\footnote{i.e.\ those DFs only depending on the
  two integrals of motion energy $E$ and the $z$-component $L_z$ of
  the angular momentum or those obeying a third integral of motion
  which is symmetric under a simultaneous change of sign of $v_r$ and
  $v_z$} (DF) the velocity ellipsoid is aligned with the coordinate
axes.  If the stellar orbits are nearly circular (i.e.\ with small
radial amplitudes with respect to the guiding centre), then the Oort ratio $X^2
\equiv \sigma^2_{\phi\phi}:\sigma^2_{rr}$ of tangential versus radial
stellar velocity dispersions can be calculated in zero order
approximation within the epicyclic theory to be $ X^2_{\rm ep}= - B / (A-B)$, where $A$ and
$B$ are the usual Oort constants. For a flat rotation curve one gets
$X^2_{\rm ep} = 1/2$. More realistic
axisymmetric distribution functions give first order corrections,
which yield values of $X^2=0.59$ (or 0.66) for the solar vicinity (Kuijken
\& Tremaine \cite{kuijken91}, hereafter KT91). Clearly, these values
exceed those predicted by the epicycle approximation. On the other hand,
observations in
the Milky Way (MW) give lower (mean) values around $X^2=0.42-0.45$, which are
substantially below the epicycle value $X^2_{\rm ep}=1/2$ (Kerr \& Lynden-Bell
\cite{kerr86}; Ratnatunga \& Upgren \cite{ratnatunga97}). The exact
reason for this discrepancy is still unknown. It might stem from a DF
asymmetric in $v_\phi$ (KT91) or from invalid assumptions, i.e.\ a
violation of stationarity or axisymmetry.

Additionally, non-vanishing vertex deviations were found in the solar vicinity in many studies (e.g.\ Wielen \cite{wielen74}, Mayor
\cite{mayor72}, Ratnatunga \& Upgren \cite{ratnatunga97}). These vertex
deviations could be explained by spiral structures (Mayor \cite{mayor70},
Yuan \cite{yuan71}), by this
supporting the importance of non-axisymmetry of the underlying
gravitational potential.  Moreover, a clear relation between $l_{\rm v}$ and
$\sigma_{rr}$ has been found. For smaller velocity dispersions larger
vertex deviations have been measured, e.g.\ for a radial velocity
dispersion of about 15 km\,s$^{-1}$ one gets a vertex deviation of
$25^\circ - 30^\circ$ (Wielen \cite{wielen74}).

A major step in analysing the local stellar velocity space was the
astrometric satellite mission Hipparcos (ESA \cite{esa97}). Earlier
results concerning the general behaviour of the velocity ellipsoid,
e.g.\ the dependence of the dispersions or the vertex deviations as a
function of $B-V$ were corroborated (Dehnen \& Binney \cite{Dehnen},
Bienaym\'e \cite{bienayme99}, Hogg et al.\ \cite{hogg05}). Moreover,
the more numerous known proper motions allowed for a detailed mapping
of the velocity space in the solar vicinity.  It turned out that the
local velocity distribution of stars exhibits a lot of substructure (Dehnen
\cite{dehnen98}, Alcob\'e \& Cubarsi \cite{alcobe05}). For example, two
major peaks in the velocity distribution were found where the smaller secondary peak is well
detached by at least 30 km\,s$^{-1}$ from the main peak (the ``u
anomaly'').  Dehnen (\cite{dehnen00}) attributed this bimodality to the
perturbation exerted by the Galactic bar assuming that its outer Lindblad
resonance (OLR) is close to the Sun. 
M\"uhlbauer \& Dehnen (\cite{muehlbauer03})
showed that a central bar can also explain vertex deviations of about
10$^\circ$, and Oort ratios below 1/2. Recent investigations about the
influence of the Galactic bar on Oort's $C$ constant corroborated
these results and gave stronger limits on the bar's properties
(Minchev et al.\ \cite{minchev07b}).
Minchev \& Quillen (\cite{minchev07a}) showed also that a {\it stationary}
spiral cannot reproduce the ''u anomaly''.
Though less likely,  an alternative interpretation of the bimodality
as a result of {\it non-stationary} spiral arms cannot be ruled out completely.

In general, a non-axisymmetric gravitational potential might lead to both, a
misalignment of the velocity ellipsoid and an additional correction of
the velocity dispersion ratio $X^2$ with respect to the value
predicted by the epicycle approximation. In the case of spiral
perturbations, this conclusion was made e.g.\ by Mayor (\cite{mayor70},
for a review of different mechanisms see KT91) and numerically
confirmed recently by Vorobyov \& Theis (\cite{VT}, hereafter VT06).
Using linear perturbation analysis KT91 studied different discs
subject to oscillating perturbations (like a rotating bar or a spiral)
or subject to a static non-axisymmetric dark matter halo.  In the case of
large-scale oscillating perturbations the predicted vertex deviations
oscillate, too.  Once the axisymmetric background gravitational potential and the
pattern speed of the perturbation are fixed, its amplitude depends in
a complex way only on the perturbed mean velocity and its spatial
gradient. For a static elliptic potential Kuijken \& Tremaine
(\cite{KT}) showed that the observed mean kinematic data ($X^2$ and
$l_{\rm v}$) can be explained, if the ellipticity of the disc is about $q
\approx 0.88$ and the Sun's location is near to the minor axis of the
corresponding potential. It is remarkable that the correction terms
introduced by the non-axisymmetry are negative (i.e.\ 
in agreement with the observations in the Milky Way), whereas the
higher order terms derived from a more realistic axisymmetric DF are
positive. Similarly, Blitz \& Spergel (\cite{blitz91}) argued for a
slowly rotating mildly triaxial Galactic potential, resulting in an
outward motion of 14 km\,s$^{-1}$ of the LSR and a vertex deviation of
10$^\circ$.

However, it is not clear if the non-axisymmetric gravitational field
is entirely responsible for the observed vertex deviation.  The
existence of moving groups of stars was shown to produce large vertex
deviations (Binney \& Merrifield \cite{BM}). The situation may become
even more complicated because moving groups of stars may in turn be
caused by the non-axisymmetric gravitational field of spiral arms.
Therefore, a detailed numerical study of the vertex deviation in
spiral galaxies is necessary.

In this paper, we present the stellar hydrodynamics simulations of linear
and non-linear stages of the spiral structure formation in stellar discs by
solving numerically the Boltzmann moment equations up to second order
in the thin-disc approximation. In contrast to linear stability analyses, we follow
the growth of the spiral structure self-consistently. Compared to $N$-body simulations,
we have the advantage that we can follow the spiral structure formation
from a very low perturbation level deeply into the non-linear
regime. The former is usually hardly accessible by 
self-consistent $N$-body simulations due to particle noise.
We study the local evolution of the velocity dispersion tensor in the plane
of the disc, 
which includes the anisotropy ($X^2$), the ratio of the smallest
versus largest principal axes of the stellar velocity ellipsoid, and the vertex
deviation ($l_{\rm v}$).  

The paper is organized as follows. In Section~\ref{method} we give a short
description of the numerical code. The model galaxy and initial conditions
are described in Section~\ref{initial}.  The physical explanation for the
growth of the spiral structure in our model disc is given in Section~\ref{spiral}.
The shape and orientation of stellar velocity ellipsoids is studied in Section~\ref{ellipse}.
In Section~\ref{discuss} we discuss the applicability of the epicycle approximation
and possible causes of the vertex deviation in spiral galaxies.
The summary and conclusions are given in Section~\ref{conclude}.

\section{The method and numerical code}
\label{method}
Stars are collisionless objects that move on orbits determined 
by the large-scale gravitational potential. An exact description of such a 
system requires the solution of the collisionless Boltzmann equation
for the distribution function of stars $f({\bl r},{\bl v},t)$ in the phase space 
($\bl{r},\bl{v}$). A general solution of the Boltzmann equation, which is defined 
in a six-dimensional position and momentum phase space, is usually prohibited due to 
the large computational expense. However, taking moments of the Boltzmann equation 
in velocity space turns out to yield a set of equations that is numerically tractable. 

In our recent paper (VT06) we presented the BEADS-2D code
that is designed to study the dynamics of stellar discs in galaxies. The BEADS-2D code
is a finite-difference numerical code that solves the Boltzmann moment equations 
up to second order in the thin-disc approximation on a polar grid ($r,\phi$). 
More specifically, the BEADS-2D code solves for the stellar
surface density $\Sigma$, mean radial and tangential stellar velocities 
$u_{r}$ and $u_{\phi}$, and stellar velocity dispersion tensor. The latter includes 
the squared radial and tangential stellar velocity dispersions $\sigma^2_{rr}$ and 
$\sigma^2_{\phi\phi}$, respectively,
and the mixed velocity dispersion $\sigma^2_{r\phi}$.
We close the system of Boltzmann moment equations by adopting the
zero-heat-flux approximation. This approximation is equivalent to assuming
the absence of any heat transfer in the classic fluid dynamics approach. 
The BEADS-2D code has shown a good agreement with the predictions of linear
stability analysis of self-gravitating stellar discs (e.g.\ Polyachenko et
al.\ \cite{Polyachenko}, Bertin et al.\ \cite{bertin00}).

The use of the Boltzmann moment equation approach allows us to study {\it non-isotropic}
stellar systems, in particular, the shape and orientation of stellar velocity ellipsoids
within the discs of spiral galaxies. The BEADS-2D code directly evolves observable quantities, 
and it can follow their evolution over many e-folding times starting from arbitrary
small perturbations. These properties make the stellar hydrodynamics approach 
advantageous over the more common N-body codes which, however, suffer strongly from particle 
noise. Presently, the BEADS-2D code is limited by 
its two-dimensionality (the thin-disc approximation) but a fully three-dimensional implementation 
is currently under development. The interested reader is referred to VT06
for details. In the current simulations, the numerical resolution has $512\times 512$ 
grid zones that are equally spaced in the $\phi$-direction and logarithmically spaced 
in the $r$-direction. The inner and outer reflecting boundaries are at 0.2~kpc and 35~kpc, 
respectively. We extend the outer computational boundary far enough to ensure a good 
radial resolution in the region of interest within the inner 20~kpc. For instance, 
the radial resolution at 1~kpc and 20~kpc is approximately 10~pc and 150~pc,
respectively.

\section{Model galaxy and initial conditions}
\label{initial}
In this paper we study the influence of a spiral gravitational field on the properties
of stellar velocity ellipsoids within the disc of a spiral galaxy. Our model galaxy
consists of a thin, self-gravitating stellar disc embedded in
a static dark matter halo. The initial surface density of stars is axisymmetric and is
distributed exponentially according to 
\begin{equation}
\Sigma(r)=\Sigma_0 \exp(-r/r_{\rm d}),
\end{equation}
with a radial scale length $r_{\rm d}$ of 4~kpc. The central surface density $\Sigma_0$
is set to $10^3~M_\odot$~pc$^{-2}$. We note that the 
densities of stellar discs are indeed found to decay exponentially with distance, 
with a characteristic scale length increasing from $2-3$~kpc for the early type galaxies 
to $4-5$~kpc in the late type galaxies (Freeman \cite{freeman}). 

The gravitational potential
$\Phi_{\rm disc}$ of the stellar disc is calculated by 
\begin{eqnarray}
  \Phi_{\rm disc}(r,\phi) & = & - G \int_0^\infty r^\prime dr^\prime 
                     \nonumber \\
      & &       \times \int_0^{2\pi} 
               \frac{\Sigma(r^\prime,\phi^\prime) d\phi^\prime}
                    {\sqrt{{r^\prime}^2 + r^2 - 2 r r^\prime
                       \cos(\phi^\prime - \phi) }}  \, .
\label{phidisc}                       
\end{eqnarray}
This sum is calculated using a FFT technique that applies the 2D Fourier 
convolution theorem for polar coordinates (Binney \& Tremaine \cite{BT}, Sect.~2.8).

\begin{figure}
   \resizebox{\hsize}{!}{
     \includegraphics[angle=0]{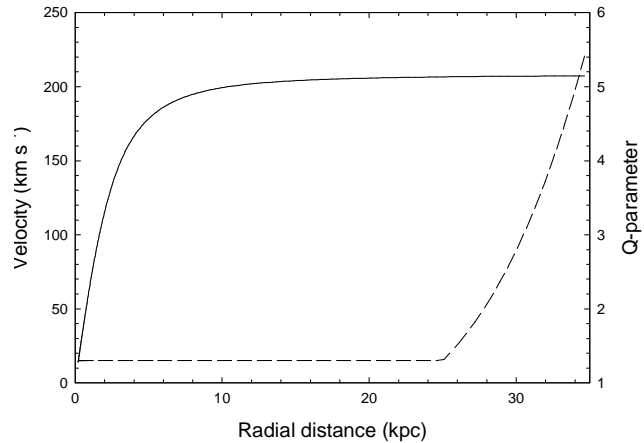}}     
   \caption{Rotation curve (solid line) and radial distribution of the Toomre $Q_{s}$ parameter
   (dashed line) in the model stellar disc. }
   \label{fig1}
\end{figure}

The initial mean rotational (tangential) velocity of stars in the disc is chosen according to
\begin{equation}
   u_{\phi} = u_\infty \cdot \left( \frac{r}{r_{\rm flat}} \right)
                 \cdot \frac{1}
                   {\displaystyle 
                    \left[ 1 + \left( \frac{r}{r_{\rm flat}} \right)^{n_v} 
                    \right]^{\displaystyle \frac{1}{n_v}}} \,\,.
   \label{vcirc}
\end{equation}
The transition radius between the inner region of rigid rotation 
and a flat rotation in the outer part is given by $r_{\rm flat}$, which
we set to 3~kpc. The smoothness of the transition is controlled by the parameter
$n_v$, set to 3. The velocity at infinity, $u_\infty$, is set to 208~km~s$^{-1}$.
The resulting rotation curve is shown in Fig.~\ref{fig1} by the solid line.
The initial mean radial velocity of the stars $u_r$ is set to zero.

The radial component of the velocity dispersion 
is obtained from the relation $\sigma_{rr}=3.36 \,Q_{\rm s}\, G \,\Sigma/\kappa$ 
for a given value of the Toomre parameter $Q_{\rm s}$. Here, $\kappa$ is the epicycle frequency. 
We assume that throughout 
most of the disc $Q_{\rm s}$ is constant and equal 1.3 but is steeply increasing 
with radius at $r>25$~kpc. The initial radial distribution of $Q_{\rm s}$ is 
shown in Fig.~\ref{fig1} by the dashed line. The tangential 
component of the velocity dispersion $\sigma_{\phi\phi}$ is determined adopting 
the epicycle approximation, in which 
the following relation between $\sigma^2_{\phi\phi}$ and $\sigma^2_{rr}$ holds 
(Binney \& Tremaine \cite{BT}, p.\ 125): 
\begin{equation} 
   \sigma_{\phi\phi}^2=\sigma_{rr}^2 \cdot {1\over2} \left({r \over u_{\rm c}} 
   {du_{\rm c} \over dr}+1 \right),
   \label{epic} 
\end{equation} 
where $u_{\rm c}$ is the circular speed, i.e. the speed of a hypothetical star in a 
circular orbit determined exclusively by the gravitational potential.
The value of $u_{\rm c}$ is yet undefined (since the gravitational potential of the dark matter halo
is yet undefined) and we substitute the circular speed with the initial rotation velocity of stars $u_\phi$.
We emphasize that equation~(\ref{epic}) is valid only for stars on {\it nearly circular} orbits.
We discuss the practical application of the epicycle approximation to spiral galaxies in
section~\ref{epicapprox}. The mixed velocity dispersion $\sigma^2_{r\phi}$ is initially set to zero.

Once the rotation curve and the radial profiles of the stellar surface density and velocity
dispersions are fixed, the dark matter halo potential $\Phi_{\rm halo}$ 
can be derived from the following steady-state momentum equation  (see eq.~(4) 
in VT06)
\begin{equation}
\Sigma \left( {\partial \Phi_{\rm disc} \over \partial r} + 
{\partial \Phi_{\rm halo} \over \partial r} \right) -
{\Sigma \sigma_{\phi\phi}^2 \over r} - {\Sigma u_\phi^2 \over r} + {1\over r}
{\partial \over \partial r} \left( r \Sigma \sigma_{rr}^2 \right)=0
\label{equilib}
\end{equation}

\section{Development of a spiral structure}
\label{spiral}


\begin{figure*}
 \centering
  \includegraphics{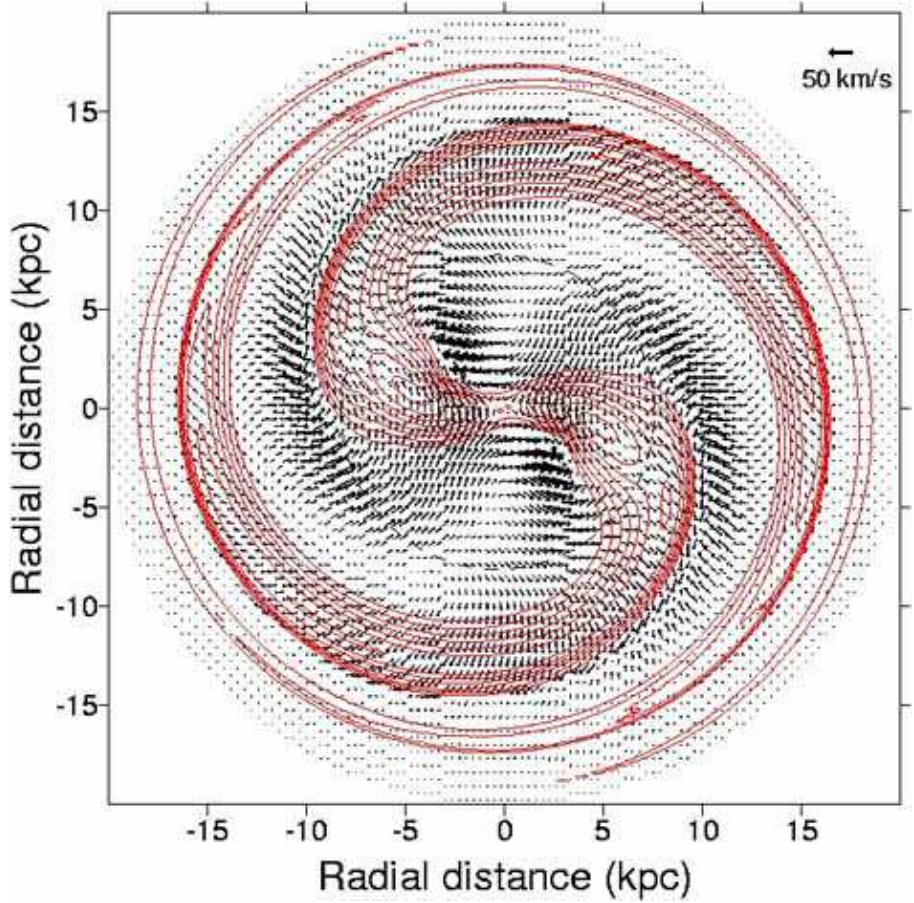}
      \caption{Contour lines of the positive stellar density perturbation relative
   to the initial axisymmetric density distribution at $t=1.6\,\, \mathrm{Gyr}$. The maximum and minimum
   contour levels are 0.8 and 0.1, respectively and the step is 0.1. The
   residual stellar velocity field in km~s$^{-1}$ is superimposed on the contour lines
   (see the text for details).}
         \label{fig2}
\end{figure*}

When starting our simulations we add a small random perturbation to 
the initially axisymmetric surface density distribution of stars. The relative 
amplitude of the perturbation in each computational zone is less than or equal
to $10^{-5}$. 
Since the model stellar disc is characterized by $Q_{\rm s}=1.3$, 
it becomes vigorously unstable to the growth of non-axisymmetric gravitational instabilities.  
We find that the positive density perturbation relative
to the initial density distribution in the disc can be a good tracer for
the spiral structure. Hence, we define the relative stellar density perturbation
as
\begin{equation}
\zeta(r,\phi) = {\Sigma(r,\phi) - \Sigma_{\rm in}(r) \over \Sigma_{\rm
in}(r) },
\end{equation}
where $\Sigma_{\rm in}(r)$ is the initial axisymmetric stellar density
distribution in the disc and $\Sigma(r,\phi)$ is the current stellar density
distribution. The positive values of $\zeta(r,\phi)$ at $t=1.6$~Gyr are  plotted in Fig.~\ref{fig2} 
by the red contour lines. The maximum and minimum contour levels correspond 
to the relative perturbations of $0.8$ and $0.1$, respectively, and the
step is $0.1$. The position of corotation
at approximately 8.5~kpc is shown by the dashed circle. 
The stellar disc has clearly developed a two-armed spiral structure
and a central bar. A movie showing the time development of the spiral
structure can be found at http://www.univie.ac.at/theis/swing\_VT07.avi.
Snapshots from different stages of the evolution have also been presented
by VT06 for a similar model stellar disc.

In our numerical simulations, the bar forms within the 4:1 ultraharmonic resonance (UHR),
which is located at approximately 4.5~kpc.
At the end of the bar, a slightly weaker spiral structure emerges
almost perpendicularly to the bar.  Such transitions of bars to
spirals are not uncommon (e.g.\ NGC 1672 or NGC 3513).
Radially, our spiral arms reach up to the OLR ($\sim$ 15 kpc).  
Azimuthally, they cover an angle of about
$\pi$ where they start to fade out.   The strongest
gradients in the stellar density are found near the outer (convex)
edges of the spiral arms. The pattern speeds of the bar and the spiral
structure are identical. Based on orbital analysis, Kaufmann \&
Contopoulos (\cite{kaufmann96}) identified
the UHR as a transition region between a bar and spiral features for
systems characterized by a single pattern speed for both the bar and the spiral.
Recently, Patsis (\cite{patsis06}) has shown that such a coupling is
related to chaotic orbits that resemble for a long time the 4:1-resonance
orbital behaviour.  We note that beyound 10~kpc our spiral arms are very
tightly wound due to the decreasing pitch angle. This is in contrast to
logarithmic spirals observed in most spiral galaxies. This inconsistency
may be caused by the presence of the OLR in
our model galaxy where spiral arms tend to converge to a resonant ring. 
We also emphasize that the amplitude of the spiral stellar density perturbation
is quickly decreasing as one approaches the OLR, which makes it difficult
to observe the stellar spirals near the OLR in real galaxies.

The non-axisymmetric gravitational
field of the spiral arms and the bar perturbs stellar orbits. To better 
illustrate this effect, we plot in Fig.~\ref{fig2} the residual velocity field of stars
at $t=1.6$~Gyr. The residual velocities are obtained by subtracting the
initial mean rotation velocities of stars (shown in Fig.~\ref{fig1}) from the
current mean velocities. Noticeable large-scale perturbations to the initially
axisymmetric flow of stars are seen between the spiral arms and near the
bar. The flow pattern near the convex edges of spiral arms (and outside
corotation) is particularly interesting. The non-axisymmetric gravitational
field of the spiral arm creates two counterpointed large-scale streams
of stars, which merge near the convex edge of the spiral into a narrow
retrograde stellar stream\footnote{We use the notion ``stellar stream'' to 
describe a large-scale non-circular motion of stars in our numerical simulations.
These streams may not be equivalent to the observed moving groups of stars, 
which are characterized by velocities and ages that are different from 
their environment. In principle, multicomponent stellar-hydrodynamics or N-body simulations are needed
to compare our stellar streams with the observed moving groups.}.
A similar phenomenon, though less apparent,
is seen near the inner edges of the bar where two merging streams of stars
produce a strongly perturbed velocity pattern. The regions with merging stellar streams 
occupy only a small portion of the stellar disc.
However, as we will see in the next section, these regions are characterised
by the most pronounced influence of the non-axisymmetric gravitational field on the shape
and orientation of stellar velocity ellipsoids.

As demonstrated recently in VT06, the most likely 
physical interpretation for the growth of a spiral structure in our model disc 
is swing amplification. Amplification occurs when any leading spiral disturbance unwinds 
into a trailing one due to differential rotation 
(Goldreich \& Lyndell-Bell \cite{GLB}, Toomre \cite{Toomre}, Athanassoula
\cite{athanassoula84}, and others). 
According to Julian \& Toomre 
(\cite{JT}), the gain of the swing amplifier in a stellar disc with $Q_{\rm s}=1.2$ and 
a flat rotation curve is the largest when $0.5\la \tilde{X} \la 2.5$. The latter quantity is defined as 
$\tilde{X} \equiv \lambda/\lambda_{\rm cr}$, where 
$\lambda \equiv 2\pi r/m $ is the circumferential wavelength of an $m$-armed spiral disturbance
and $\lambda_{\rm cr} \equiv 4\pi^2 G \Sigma/\kappa^2$ is the longest unstable wavelength 
in a cold disc. 
Generally speaking, swing amplification is most efficient when 
$\lambda \approx \lambda_{\rm cr}$. At $\lambda \gg \lambda_{\rm cr}$ and $\lambda \ll \lambda_{\rm cr}$, 
the swing amplifier is strongly moderated by local shear and random motion of stars, respectively.

\begin{figure}
   \resizebox{\hsize}{!}{
     \includegraphics[angle=0]{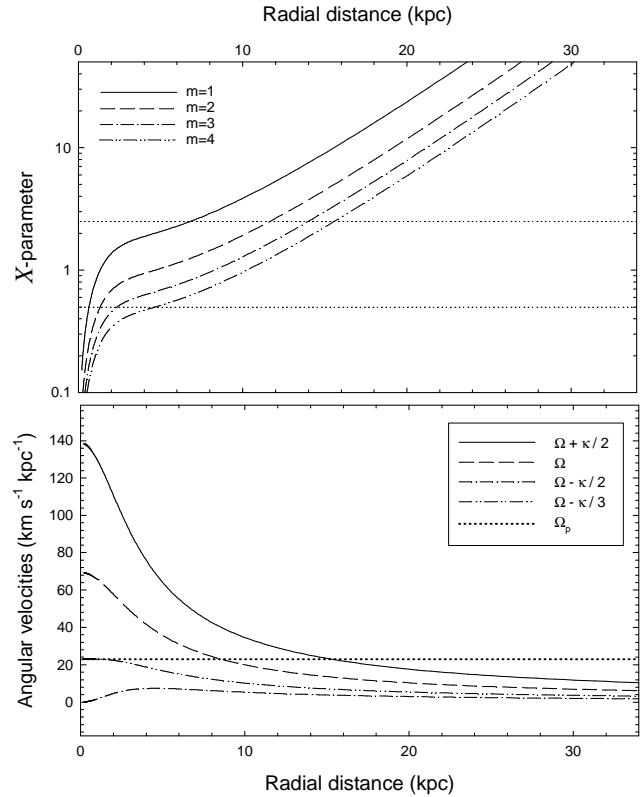}}     
   \caption{{\bf Top panel}. Radial profiles of the $\tilde{X}$-parameter for the $m=1$ (solid line),
$m=2$ (dashed line), $m=3$ (dash-dotted line), and $m=4$ (dash-dot-dotted line) 
spiral perturbations. Two horizontal dotted lines delimit the region of maximum swing amplification.
{\bf Bottom panel.}  Radial profiles of the angular velocity of stars $\Omega$ and $\Omega
\pm \kappa/m$  for the initial axisymmetric stellar disc and two spiral modes $m=2$, and $m=3$.
 The angular velocity of the spiral pattern $\Omega_{\rm p}$ is plotted
 by the dotted line. There is no inner Lindblad resonance for the $m=2$ mode,
 whereas the $m=3$ mode has the inner Lindblad resonance near the disc
 centre.}
   \label{fig3}
\end{figure}

The top panel of Fig.~\ref{fig3} shows the radial profiles of the $\tilde{X}$-parameter for the $m=1$ (solid line),
$m=2$ (dashed line), $m=3$ (dash-dotted line), and $m=4$ (dash-dot-dotted line) 
spiral perturbations. The region of maximum swing amplification is 
bound by the two horizontal dotted lines. 
It can be clearly seen that the $m=1$ perturbations are swing amplified in the inner
5~kpc, whereas the $m=2$ perturbations are swing amplified over a much larger area
between 1~kpc and 12~kpc. Consequently, the $m=2$ perturbations are expected to engage 
more stellar mass into the amplification process than the $m=1$ perturbations. 
Hence, it is natural to conclude that the $m=2$ density perturbations should grow 
faster and ultimately surpass the $m=1$ perturbations. 

Why then does our model stellar disc develop a two-armed global spiral pattern rather than
a three-armed  or any other multi-armed pattern? Indeed, higher order perturbations are also 
swing amplified over a considerable area of the disc, as Fig.~\ref{fig3}
(top panel) demonstrates. The answer is that swing amplification 
needs a feedback mechanism that can constantly feed a disc with leading spiral disturbances.
The trailing short-wavelength disturbances may propagate through the centre and emerge 
on the other side as leading ones, thus providing a feedback for the swing amplifier. 
Figure~4 in VT06 and the mentioned movie nicely illustrate this phenomenon.
However, the propagation of stellar disturbances through the disc centre is possible if there is 
no inner Lindblad resonance (ILR) or if the waves are reflected before reaching the ILR.  
A $Q$-barrier allows for such a reflection of propagating waves (e.g.\ Athanassoula \cite{athanassoula84}) 
but it is absent in the inner region of our model disc 
(see Fig.~\ref{fig1}).

To determine the position of Lindblad resonances $m(\Omega_{\rm p} -\Omega)=\pm \kappa$,
we plot in the bottom panel of Fig.~\ref{fig3} the radial profiles of $\Omega$ and $\Omega\pm \kappa/m$  
for the initial axisymmetric stellar disc and two global spiral modes $m=2$ and $m=3$. 
The angular velocity of the 
two-armed global spiral pattern (see next section) is determined from both visual
and numerical estimates to be $\Omega_{\rm p}=23 \pm 1$~km~s$^{-1}$~kpc$^{-1}$.
This value is plotted by the dotted line in the bottom panel of Fig.~\ref{fig3}. It is clearly seen that
the $m=2$ disturbances have no inner Lindblad resonance, $\Omega-\kappa/2 <\Omega_{\rm p}$ at 
all radii. On the other hand, the $m=3$ disturbances have the inner Lindblad resonance
near the disc centre, $\Omega -\kappa/3 \approx \Omega_{\rm p}$ in the inner 2~kpc. 
This  prevents the $m=3$ (and any higher order) disturbances from propagating through 
the disc centre, which terminates the feedback mechanism for the swing amplifier for
these modes. We find that this mechanism works nicely in stellar discs with 
other radial configurations of the $\tilde{X}$-parameter and pattern angular velocities $\Omega_{\rm p}$
and can allow for the dominant growth of $m=1$, $m=3$, and other higher order global spiral 
modes. The results of this study will be presented in a future paper.

\section{The shape and orientation of stellar velocity ellipsoids}
\label{ellipse}

\subsection{Stellar velocity dispersions and spiral structure}
\label{epicapprox}

 Despite more exotic means of galactic disc heating there are 
currently two known mechanisms involving spirals as 
the heating agents. These are transient spiral stellar density waves
(e.g. Wielen \cite{Wielen}, Carlberg \& Sellwood \cite{CS}, Jenkins \&
Binney \cite{JB}, De Simone et al. \cite{DeSimone}) and multiple spiral
stellar density waves (Minchev \& Quillen \cite{Minchev06}). 
The latter mechanism involves two sets of steady state spiral structure
moving at different pattern speeds.
We note that a single steady state spiral structure as defined in e.g. Lin et
al. (\cite{Lin}) does not heat stellar discs. 

\begin{figure}
   \resizebox{\hsize}{!}{
     \includegraphics[angle=0]{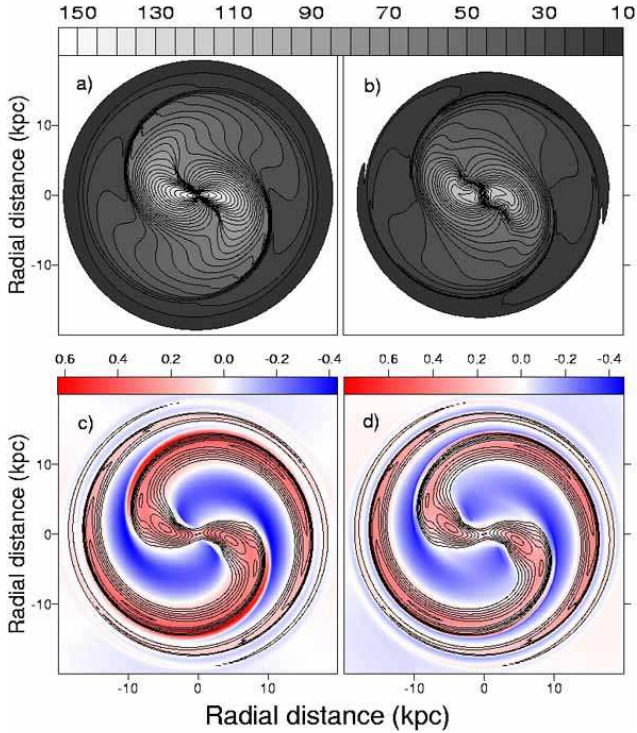}}     
   \caption{Spatial distribution of ({\bf a})  the radial and ({\bf b}) tangential
   stellar velocity dispersions at $t=1.6$~Gyr. The scale bar is in km~s$^{-1}$.
   The spiral structure is clearly seen in the distribution of both velocity
   dispersions.
   Perturbations in
  ({\bf c}) the radial  and  ({\bf d}) tangential velocity dispersions  at $t=1.6$~Gyr 
   relative to their initial values. The contour lines of the positive stellar
   density perturbations are shown in ({\bf c}) and ({\bf d}) for convenience (see Fig.~\ref{fig2}
   for details). Note that the positive relative perturbations
   in the velocity dispersions correlate spatially with the positive relative
   density perturbations shown in Fig.~\ref{fig2}. }
   \label{fig4}
\end{figure}

In this section, we focus on the effect that the non-axisymmetric gravitational 
field of a swing amplified spiral structure may have on the spatial
distribution of stellar velocity dispersions within a stellar disc. 
Figures~\ref{fig4}a and \ref{fig4}b show the spatial distribution of the radial and tangential
stellar velocity dispersions, respectively, at $t=1.6$~Gyr. A two-armed spiral structure
and a bar are apparent in the spatial distribution of $\sigma_{rr}$ and
$\sigma_{\phi\phi}$. To better illustrate this phenomenon, we plot in Figs~\ref{fig4}c
and \ref{fig4}d the stellar velocity dispersion perturbations relative to their initial values at $t=0$~Gyr. 
More specifically, Fig.~\ref{fig4}c shows the relative perturbations in
the radial velocity dispersion calculated as $(\sigma_{rr}-\sigma^0_{rr})/\sigma^0_{rr}$, where $\sigma^0_{rr}$ 
is the initial radial velocity dispersion and $\sigma_{rr}$ is the model's known radial
velocity dispersion at $t=1.6$~Gyr.  To avoid confusions, we note that
$\sigma_{rr}$ and $\sigma_{\phi\phi}$ are measured in km~s$^{-1}$ throughout
the paper. Figure~\ref{fig4}d shows 
the relative perturbations in the tangential velocity dispersion $\sigma_{\phi\phi}$ calculated
in a similar manner. 
The positive and negative perturbations are plotted with the shadows of red and blue, respectively.
The contour lines of the positive stellar density perturbations are shown
for convenience. Figure~\ref{fig4} clearly demonstrates that the non-axisymmetric 
gravitational field has a pronounced 
global influence on the spatial distribution of stellar velocity dispersions in the disc.
The most distinctive feature in Fig.~\ref{fig4} is that the positive (and
negative) relative
perturbations in density and both stellar velocity dispersions correlate spatially,
suggesting a causal link. We emphasize that the largest spatial gradients
in both velocity dispersions (and in the stellar surface density) are found
near the outer (convex) edges of the spiral arms. As we will see in Section~\ref{epicycle},
this may cause a severe failure of the epicycle approximation in these
regions.

\subsection{The shape of stellar velocity ellipsoids}
\label{ellipsoid}

In contrast to gaseous discs, stellar discs are essentially non-isotropic.
This means that the local properties of stars, such as mean velocities, 
are determined by the velocity dispersion 
tensor $\bl \sigma$ rather than by an isotropic pressure. The tensor $\bl \sigma$
is often non-diagonal in the local coordinate system ($r,\phi,z$).
The principal axes of a diagonalized velocity dispersion tensor form an imaginary ellipsoidal 
surface that is called the velocity ellipsoid.
The available measurements in the solar vicinity
indicate that the ratio $\sigma_1:\sigma_2$ of the smallest versus largest principal 
axes of the stellar velocity ellipsoid in the disc plane does not vary significantly 
with the $B-V$ colour.  According to Dehnen \& Binney (\cite{Dehnen}), 
the axis ratio $\sigma_1:\sigma_2 $ is approximately $0.6\pm 0.1$  for stars with 
$-0.2 \le B-V \le 0.7$. To avoid confusions, we note that throughout the paper $\sigma_1$
and $\sigma_2$ are measured in km~s$^{-1}$. The $B-V$ colour of an ensemble of stars may be considered 
as a rough estimate of their mean age. A weak sensitivity of the axis ratio
$\sigma_1:\sigma_2$  to the age of stellar populations is intriguing since 
the principal axes of velocity ellipsoids (when considered separately) are indeed age sensitive and 
show a noticeable (a factor of two) increase along increasing $B-V$ colours 
(Dehnen \& Binney \cite{Dehnen}).

\begin{figure}
   \resizebox{\hsize}{!}{
     \includegraphics[angle=0]{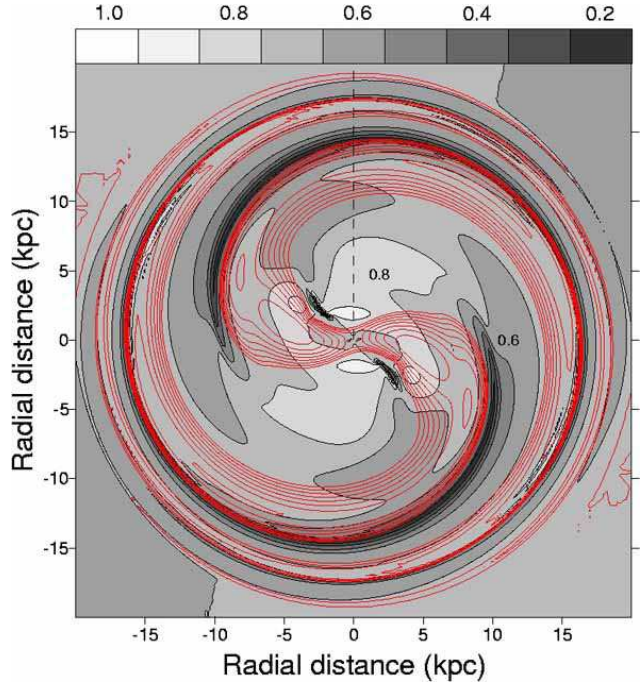}}     
   \caption{Contour map of the positive stellar density perturbations (red
   lines) superimposed
   on the ratio $\sigma_1:\sigma_2$ of smallest versus largest principal
   axes of stellar velocity ellipsoids at $t=1.6$~Gyr. Note abnormally
   small values of $\sigma_1:\sigma_2$ at the outer (convex) edges of spiral arms.}
   \label{fig5}
\end{figure}

\begin{figure}
   \resizebox{\hsize}{!}{
     \includegraphics[angle=0]{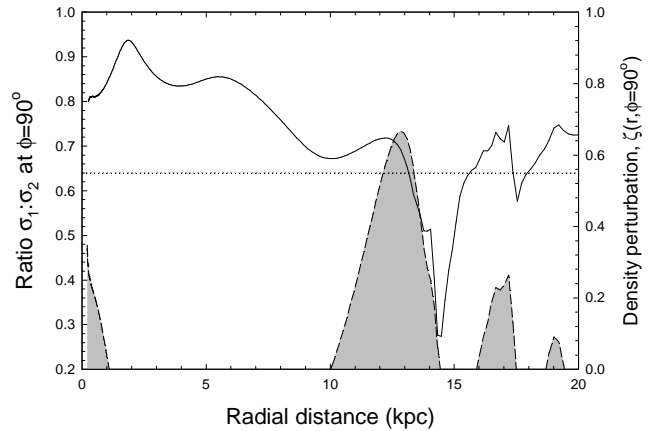}}     
   \caption{Radial profiles of the ratio $\sigma_1:\sigma_2$ (solid line)
   and positive relative perturbation in the stellar surface density (dashed
   line) obtained from Fig.~\ref{fig5} by taking a radial cut at the
   azimuthal angle $\phi=90^\circ$. Note that $\sigma_1:\sigma_2$ attains
   local minima just at the outer (convex) edges of spiral arms.}
   \label{fig6}
\end{figure}

If the axis ratio $\sigma_1:\sigma_2$ is indeed only 
weakly sensitive to the age of stellar populations, it would be interesting to 
determine other phenomena that could possibly affect the values of $\sigma_1:\sigma_2$.
In this section, we consider the effect that the non-axisymmetric gravitational
field of the spiral arms and the bar may have on the shape of stellar velocity
ellipsoids within the disc of our model galaxy. Figure~\ref{fig5} shows the contour plot of 
the positive stellar density perturbation (red lines) superimposed on the
grey-scale map of $\sigma_1:\sigma_2$ in the disc at $t=1.6$ Gyr. It is evident that
the central 5~kpc are characterized by $\sigma_1:\sigma_2 \approx 0.8-0.9$.
Most of the intermediate and outer disc regions have $\sigma_1:\sigma_2 
\approx 0.6-0.7$, which is comparable to the solar neighbourhood value of
0.64 (Dehnen \& Binney \cite{Dehnen}). However, 
parts of the stellar disc that are immediately adjacent to the outer (convex) edges of 
spiral arms are characterized by a noticeably smaller value of 
$\sigma_1:\sigma_2 \approx 0.25-0.5$. To better quantify this phenomenon,
we calculate the axis ratio $\sigma_1:\sigma_2$ along a radial cut at the azimuthal
angle $\phi=90^\circ$, which is
shown in Fig.~\ref{fig5} by the dashed line. The resulting ratio $\sigma_1:\sigma_2$
is plotted in Fig.~\ref{fig6} by the solid line, whereas the dashed line
shows the relative stellar density perturbation $\zeta(r,\phi)$ 
obtained along the same radial cut.  The area below the dashed line is  filled
with gray to better identify the position of spiral arms. The dotted line represents
the solar neighbourhood value $\sigma_1:\sigma_2\approx 0.64$. It is evident that
the axis ratio $\sigma_1:\sigma_2$ has deep minima near the outer (convex) edges
of the spiral arms. The smallest value of $\sigma_1:\sigma_2\approx 0.25$
is attained at $r\approx 14.5$~kpc. 

The cause for such small values of $\sigma_1:\sigma_2$ near the outer
edges of spiral arms is unclear.  One of the principal 
velocity dispersions $\sigma_1$ or $\sigma_2$ becomes much  smaller than the other, if the 
mixed velocity dispersion $\sigma^2_{r\phi}$ becomes comparable to
the squared radial ($\sigma^2_{rr}$) and tangential ($\sigma^2_{\phi\phi}$) 
velocity dispersions. We will see in the next section that the outer (convex) edges of
the spiral arms are indeed characterized by large values of $\sigma^2_{r\phi}$ and
associated vertex deviations. These large values of the mixed velocity dispersion
are most likely caused by  the two large-scale (counterpointed) streams of stars that merge near the 
outer edge of the spiral arms into a retrograde narrow stellar stream (see Fig.~\ref{fig2}).
It is not clear that such streams are a general property of spiral galaxies and
high-resolution observation are necessary to confirm their existence.

The (possible) existence of regions with abnormally low ratios of $\sigma_1:\sigma_2$
can be potentially used to determine the position of {\it stellar} spiral arms. 
This task is not trivial since the spiral arms in external galaxies are usually manifested
by strong emission from current star formation, which traces the gaseous response to the
underlying stellar spiral arms. There could exist a substantial phase shift between
the gaseous and stellar spiral arms. Figure~\ref{fig5} indicates that in the inter-arm region 
$\sigma_1:\sigma_2 \approx 0.6-0.7$,
while near the convex edge of the arm $\sigma_1:\sigma_2$ can become as low as 0.25.
Such a large difference can be detected. Moreover, our numerical simulations show that
a larger contrast is expected for gravitationally more unstable discs. 
For instance, in a $Q_{\rm s}=1.1$ stellar disc the $\sigma_1:\sigma_2$ ratio
near the convex edges of spiral arms can become as small as $0.15$ but
it stays near $0.6-0.7$ in the inter-arm region. 
On the other hand, multi-armed spirals usually show a much smaller contrast
in the $\sigma_1:\sigma_2$ ratio, possibly due to smaller perturbations
to the velocity field of stars. 
High-resolution measurements of the stellar velocity ellipsoids in grand-design
two-armed spiral galaxies are needed to confirm the existence of abnormally 
low values of the axis ratio $\sigma_1:\sigma_2$.

  Finally, it should be stressed that our models are genuine
  2D-models.  From a higher-order moment analysis Cuddeford \& Binney
  (\cite{cuddeford94}) showed that neglecting the $z$-motion can lead
  to a systematic overestimate of $\sigma_{\phi\phi}$ and, thus, an
  overestimate of the Oort ratio $\sigma^2_{\phi\phi} /
  \sigma^2_{rr}$.  Though such 3D-effects will definitely have an
  impact on the exact Oort ratio, the changes are about 20-25\% (cf.\
  e.g.\ fig.\ 2 in Cuddeford \& Binney).  Therefore, we do not expect
  them to alter our results qualitatively, i.e.\ the large spatial
  variations of $\sigma_1:\sigma_2$ found in our analysis will be
  kept. In fact, if $\sigma^2_{\phi\phi}$ is smaller than
  $\sigma^2_{rr}$, the ratio $\sigma^2_{\phi\phi} / \sigma^2_{rr}$
  should be overestimated in our 2D-models and the mentioned variation
  should be even larger.

\subsection{Vertex deviation}
\label{sect_vertex}

\begin{figure}
   \resizebox{\hsize}{!}{
     \includegraphics[angle=0]{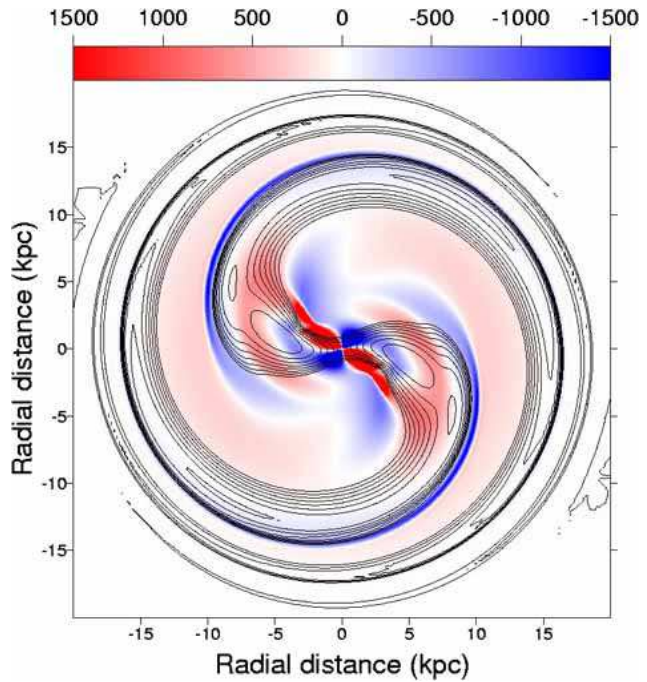}}     
   \caption{Contour map of the positive stellar density perturbation superimposed
   onto the map of the mixed velocity dispersion $\sigma^2_{r\phi}$. The scale bar is
   in $({\rm km~s}^{-1})^2$. }
   \label{fig7}
\end{figure}

It has long been known from observations that the mixed velocity dispersion
within the disc plane is non-zero for stars of many stellar types
(see e.g. Dehnen \& Binney \cite{Dehnen}). The BEADS-2D code directly evolves
the mixed velocity dispersion $\sigma^2_{r\phi}$ and is particularly
suitable for studying the evolution of $\sigma^2_{r\phi}$ in galactic discs.
Figure~\ref{fig7} shows the map of $\sigma^2_{r\phi}$ at $t=1.6$~Gyr. 
Positive and negative values of $\sigma^2_{r\phi}$ are shown with the shadows of red and
of blue, respectively. We emphasize the fact that $\sigma^2_{r\phi}$ 
may be both negative and positive. Only the values between $-1500~({\rm km~s}^{-1})^2$ and 
$+1500~({\rm km~s}^{-1})^2$ are plotted to emphasize $\sigma^2_{r\phi}$ 
in the outer disc. It is evident that $\sigma^2_{r\phi}$ takes large 
values near the convex edges of spiral arms and in the inner central region.  
For instance, $\sigma^2_{r\phi}$ can become as large as
$-1500~({\rm km~s}^{-1})^2$ near the convex edges of spiral arms and
$\pm 5000~({\rm km~s}^{-1})^2$ near the disc centre.
The presence of substantially non-zero $\sigma^2_{r\phi}$ implies
that the axis ratio $\sigma_1:\sigma_2$ may become quite different from the ratio 
$\sigma_{\phi\phi}:\sigma_{rr}$ of the tangential versus radial velocity dispersions.

The non-zero $\sigma^2_{r\phi}$ 
indicate that the stellar velocity ellipsoids are not aligned along the local coordinate
vector $\hat{r}$ (galactic centre-anticentre direction). The degree of this misalignment 
is often quantified in terms of a vertex deviation.  A classic definition for the 
vertex deviation is 
(Binney \& Merrifield \cite{BM}, p. 630)
\begin{equation}
l_{\rm v}={1 \over 2} {\rm atan} \left( {2 \sigma^2_{r\phi} \over \sigma^2_{rr}-
\sigma^2_{\phi\phi}}   \right).
\label{vertex}
\end{equation}
According to this definition, the vertex deviation takes the values between 
$-45^\circ$ and $+45^\circ$. In practice, however, the degree of misalignment
of the major axis of stellar velocity ellipsoids with the centre-anticentre direction 
may exceed these bounds, especially
in the regions where the epicycle approximation breaks down.
In an extreme case, one might consider a velocity distribution with a vanishing
non-diagonal element of $\bl{\mathbf{\sigma}}$ and $\sigma_{\phi\phi}$ exceeding $\sigma_{rr}$. 
Then the major principal axis of the velocity ellipsoid will be aligned with the azimuthal 
direction, yielding $l_{\rm v} = \pi/2$.
Therefore, we have extended the classic definition (\ref{vertex}) as 
in VT06
\[ \tilde{l}_{\rm v} = \left\{ \begin{array}{ll} l_{\rm v}
    & \,\, \mbox{if $\sigma^2_{rr} > \sigma^2_{\phi\phi}$} \\
   l_{\rm v} + {\rm sign}\left(\sigma^2_{r\phi}\right) \cdot \displaystyle\frac{\pi}{2} 
    & \,\, \mbox{if $\sigma^2_{rr}<\sigma^2_{\phi\phi}$},   \end{array}
   \right. \]
According to this re-definition, the vertex deviation can take the values
between $-90^\circ$ and $+90^\circ$. 
In this paper we extend our previous analysis (VT06) and search for a correlation between 
the magnitude (and sign) of the vertex deviation and the characteristics
of the spiral gravitational field. 

Global Fourier amplitudes are often used to quantify the growth rate of
spiral perturbations in gravitationally unstable stellar and gaseous discs.
These amplitudes are defined as
\begin{equation}
A_{\rm m}(t)= {1 \over M_{\rm disc}} \left| \int^{r_{\rm out}}_{r_{\rm
in}} \int^{2\pi}_{0} \Sigma(r,\phi,t) e^{i m \phi} \, r\, dr\, d\phi
\right|,
\end{equation}
where $m$ is the spiral mode, $M_{\rm disc}$ is the total disc mass, 
and $r_{\rm in}$ and $r_{\rm out}$ are the disc inner and outer radii, 
respectively. The global Fourier 
amplitudes $A(m)$ provide  a rough estimate on the total relative amplitude of 
a spiral density perturbation with mode $m$. 

\begin{figure}
   \resizebox{\hsize}{!}{
     \includegraphics[angle=0]{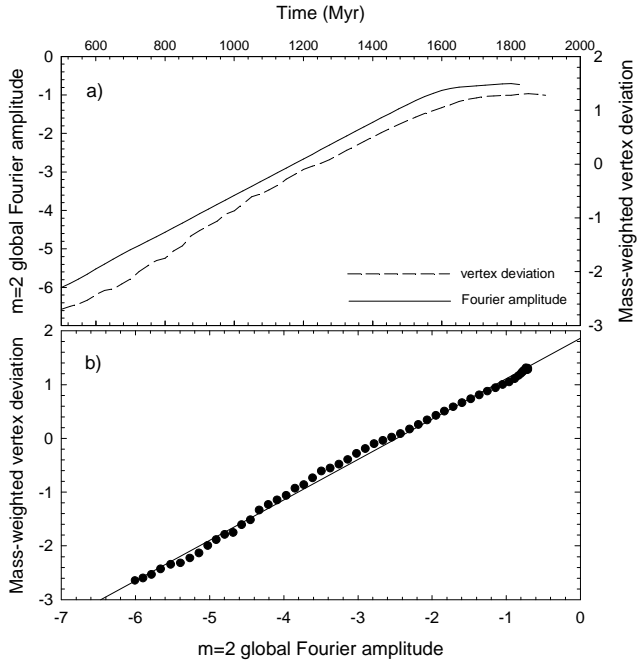}}     
   \caption{({\bf a}) Temporal evolution of the $m=2$ global Fourier amplitude
   $A_2$ (solid line) and mass-weighted vertex deviation $\bar{l}_{\rm v}$ (dashed line) in our
   model stellar disc. ({\bf b}) Dependence of $\bar{l}_{\rm
   v}$ on $A_2$ (filled circles). The least-square fit (solid line) yields a near-linear
   relation $\bar{l}_{\rm v} \propto A_2^{0.75}$. }
   \label{fig8}
\end{figure}

The solid line in Fig.~\ref{fig8}a shows the temporal evolution of the 
dominant $m=2$ global Fourier amplitude (in log units) in our model stellar disc. 
The $m=2$ amplitude indicates a steady growth of a two-armed spiral structure 
in the disc from 0.5~Gyr till approximately 1.6~Gyr, when it saturates at $\log A_2(t)\approx -0.8$. 
The subsequent evolution of 
the stellar disc is characterized by a near-constant value of $A_{2}(t)$ for 
approximately another 300~Myr, during which the spiral structure is 
being destroyed by the gravitational torques from spiral arms (this phenomenon is discussed
in detail in VT06). It is interesting to see the temporal behaviour 
of the vertex deviation as the spiral structure grows, saturates, and disperses.
The dashed line in Fig.~\ref{fig8}a shows the temporal evolution of the mass-weighted
vertex deviation (in log units) defined as
\begin{equation}
\bar{l}_{\rm v}=  M_{\rm disc}^{-1} \, \sum\limits_i  dm_{\rm i} \, \tilde{l}_{\rm v,i}
\,\, ,
\end{equation}
where $dm_{\rm i}$ is the stellar mass in the $i$-th computational zone, 
$\tilde{l}_{\rm v,i}$ is the vertex deviation in the $i$-th zone, and summation is performed
over all computational zones. We plot the mass-weighted vertex deviation
rather than its arithmetic average to deemphasize high vertex deviations 
in regions of low stellar density. A correlation between $A_2(t)$
and $\bar{l}_{\rm v}$ is clearly evident -- both grow and saturate 
in a similar manner. The saturated value of the mass-weighted vertex deviation is
approximately $15^\circ$. To emphasize the correlation between $A_2$ and $\bar{l}_{\rm
v}$,  we plot in Fig.~\ref{fig8}b the mass-weighted vertex deviation versus the $m=2$ global 
Fourier amplitude (filled circles), both in log units.
The least-square fit shown by the solid lines yields a near-linear relation, $\bar{l}_{\rm v}\propto A_2^{0.75}$.

\begin{figure}
   \resizebox{\hsize}{!}{
     \includegraphics[angle=0]{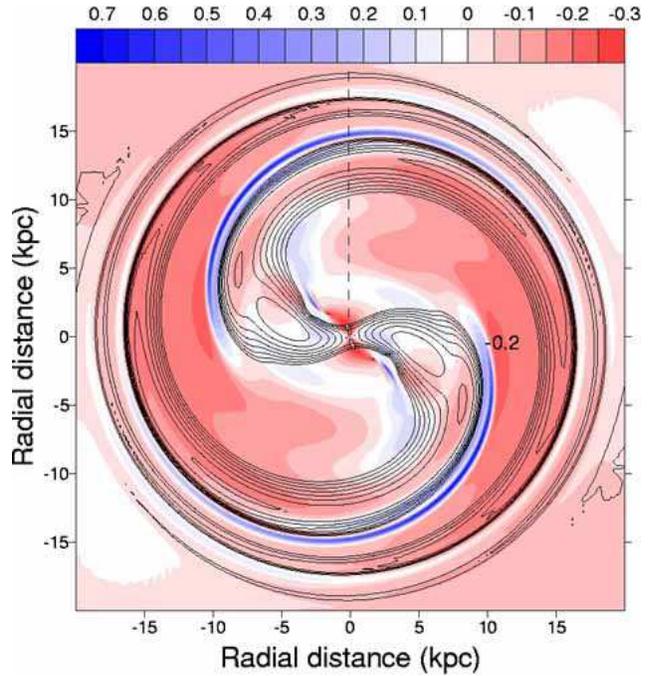}}     
   \caption{Positive stellar density perturbation superimposed on the vertex
   deviation map at $t=1.6$~Gyr. Positive and negative vertex deviations
   are shown with the shadows of red and blue, respectively. The numbers indicate the
   maximum positive vertex deviation in the inter-arm region and maximum negative 
   vertex deviation in the outer disc. The scale
   bar is in degrees.}
   \label{fig9}
\end{figure}

While the mass-weighted vertex deviation appears to correlate {\it globally} with 
the amplitude of spiral density perturbations, 
it is not clear that this correlation should hold {\it locally} as well. 
Indeed, Fig.~\ref{fig9} shows the vertex deviations in the disc at $t=1.6$~Gyr.
The positive and negative vertex deviations are plotted with the shadows of red and 
blue, respectively. White space corresponds to a near-zero vertex deviation.
The black contour lines in Fig.~\ref{fig9} delineate the spiral pattern by showing 
the regions with a positive relative stellar density perturbation $\zeta(r,\phi)$. 
The minimum and maximum contour 
levels correspond to $\zeta=0.1$ and $\zeta=0.8$, respectively. 
It is clearly seen that the value (and sign) of the vertex deviation 
{\it does not correlate} with the amplitude of the positive stellar density perturbation.
In fact, regions with maximum amplitude are characterized by 
near-zero or small negative vertex deviations ($\tilde{l}_{\rm v} \le -5^\circ$). 
At the same time,
the convex edges of spiral arms, where the stellar density perturbations are small 
but the gradients of the density perturbations are large, are characterized by
quite large negative vertex deviations reaching $\tilde{l}_{\rm v}=-40^\circ$ 
in the outer disc. Considerable positive vertex deviations up to $+30^\circ$ 
are seen between the spiral arms, where the stellar density perturbations are negative.
Somewhat surprisingly, the bar has a near-zero vertex deviation but
the central regions on both sides of the bar are characterized by 
the largest vertex deviations up to $\tilde{l}_{\rm v} = \pm 90^\circ$.
These large values of $\tilde{l}_{\rm v}$ are off the scale in Fig.~\ref{fig9}
(and subsequent figures) to emphasize the regions with lower vertex deviations\footnote{We note that the sign of the vertex deviation changes when the galactic rotation is reversed
from the counterclockwise rotation (as in present numerical simulations) to the 
clockwise rotation. This happens because the mean tangential velocities ($\bar{v}_\phi \equiv u_\phi$) 
and instantaneous  tangential velocities ($v_\phi$) of stars change the sign but 
the mean and instantaneous radial velocities ($\bar{v}_r \equiv u_r$ and $v_r$, respectively) 
do not. As a consequence, the mixed velocity dispersion
defined by equation~(\ref{mixed_vd}) and the corresponding vertex deviations 
change the sign, too.}.


\section{Discussion}
\label{discuss}


\subsection{Comparison with the epicycle approximation}
\label{epicycle}
It is often observationally difficult to measure both radial and tangential velocity
dispersions of stars. Hence, it is tempting to use equation~(\ref{epic}) to derive 
either of the two dispersions from the known one and the circular speed of stars. 
One has to keep in mind that equation~(\ref{epic}) is valid only in the epicycle approximation, 
when the epicycle amplitude (deviation from a purely circular orbit) at a 
specific radial distance $r$ is much less than both $r$ and the scale length for changes 
in both, the stellar density and dispersion (Kuijken \& Tremaine \cite{KT}, Dehnen \cite{dehnen99}). 
The latter condition may become quite stringent, especially near the convex edges of spiral arms.
Therefore, it is important to know if equation~(\ref{epic})
holds in spiral galaxies with a well-developed spiral structure.

To test the validity of equation~(\ref{epic}), we calculate the ratio 
$\sigma_{\phi\phi}:\sigma_{rr}$ from the model's known velocity dispersions and compare it with 
the ratio 
$(\sigma_{\phi\phi}:\sigma_{rr})_{\rm ep}$ derived using equation~(\ref{epic}).
From the observational point of view, it is usually difficult to know the exact value of 
the circular speed $u_{\rm c}$ that enters equation~(\ref{epic}), 
since it requires independent measurements of the axisymmetric gravitational potential in a galaxy.  
The observers usually apply the {\it local} and azimuthally averaged rotation velocities 
of stars $u_\phi$ and $\bar{u}_\phi$, respectively, as a proxy for $u_{\rm c}$.
The validity of this assumption is checked below.

\begin{figure}
   \resizebox{\hsize}{!}{
     \includegraphics[angle=0]{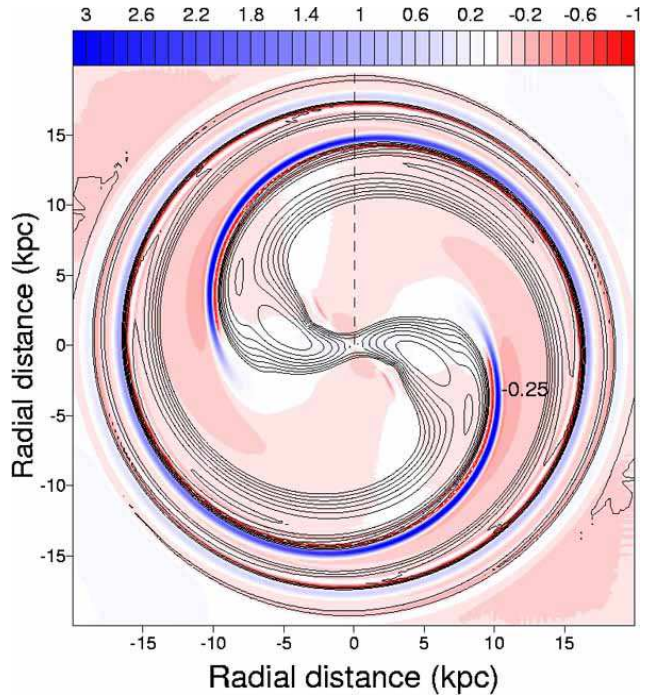}}     
   \caption{Relative errors between the ratio $(\sigma_{\phi\phi}:\sigma_{rr})$
   determined from the model's known velocity dispersions
   and ratio $(\sigma_{\phi\phi}:\sigma_{rr})_{\rm ep}$ determined from the epicycle 
   approximation~(\ref{epic}) using the local rotational velocity $u_{\phi}$
   as a proxy for the circular speed $u_{\rm c}$. The positive and negative relative errors are 
   shown with the shadows of blue and red, respectively. The black contour lines delineate the
   positive stellar density perturbation at $t=1.6$~Myr.  }
   \label{fig10}
\end{figure}

\begin{figure}
   \resizebox{\hsize}{!}{
     \includegraphics[angle=0]{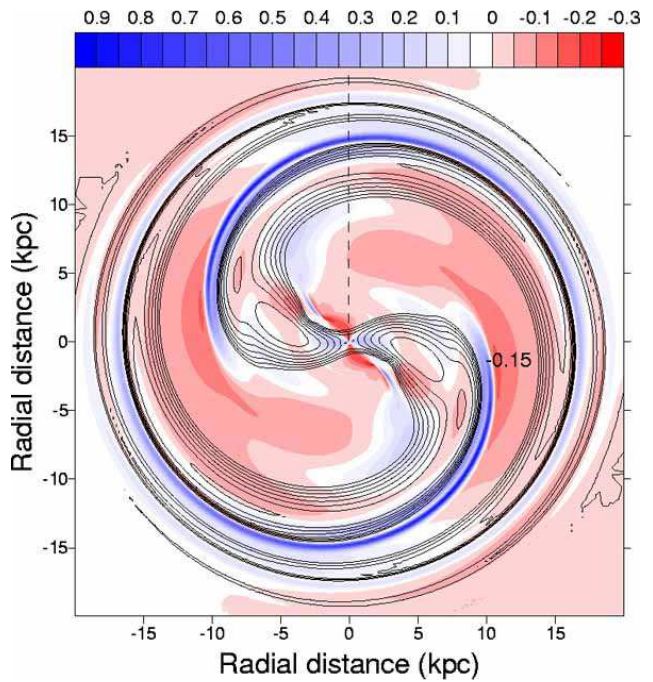}}     
   \caption{Same as Fig.~\ref{fig10} only with the azimuthally averaged rotational velocity $\bar{u}_{\phi}$
   used as a proxy for the circular speed $u_{\rm c}$.}
   \label{fig11}
\end{figure}

To quantify the deviations from the epicycle approximation in our model spiral galaxy,
we calculate the relative errors 
\begin{equation} 
\xi(r,\phi)={ (\sigma_{\phi\phi}:\sigma_{rr})_{\rm ep} - \sigma_{\phi\phi}:\sigma_{rr} 
\over \sigma_{\phi\phi}:\sigma_{rr} }
\end{equation}
using different approximations for the circular speed $u_{\rm c}$. 
In particular, Fig.~\ref{fig10} and Fig.~\ref{fig11} show
the relative errors that are obtained assuming the local rotational velocity $u_\phi$
and azimuthally averaged rotational velocity $\bar{u}_\phi$ as a substitute for
the circular speed $u_{\rm c}$, respectively. The positive and negative errors
are plotted with the shadows of blue and red, respectively.
The spiral pattern in the disc is outlined by contour lines showing 
the positive relative stellar density perturbations at $t=1.6$~Gyr.

\begin{figure}
   \resizebox{\hsize}{!}{
     \includegraphics[angle=0]{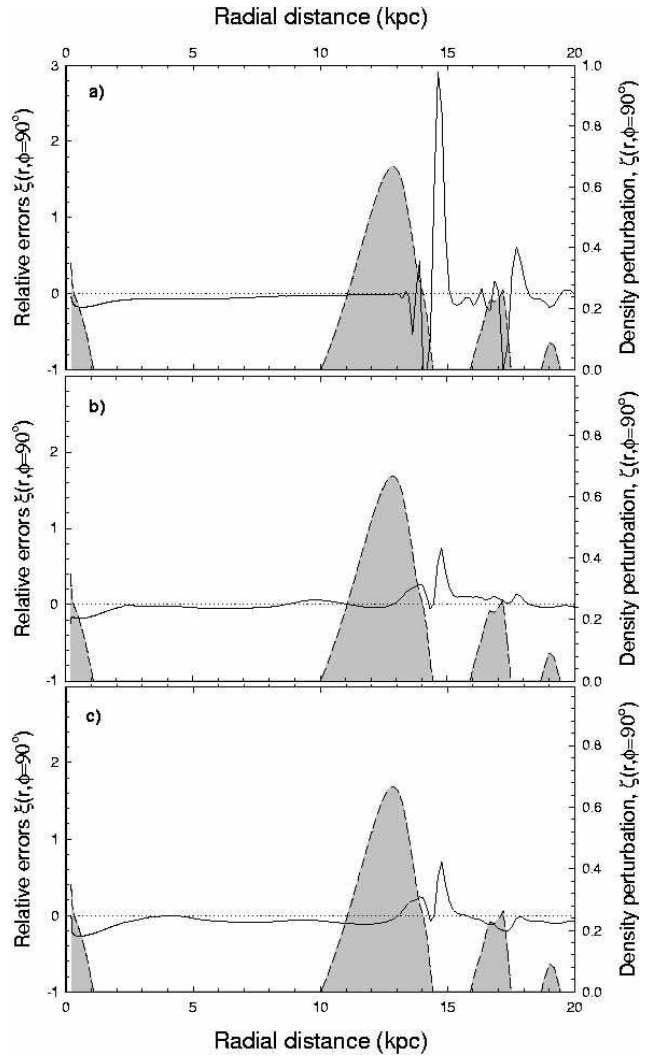}}     
   \caption{Radial profiles of the relative error $\xi(r,\phi=90^\circ)$ (solid lines)
   obtained from {(\bf a)} Figure~\ref{fig10}, {(\bf b)} Figure~\ref{fig11}, and
   {(\bf c)} Figure~\ref{fig13} by taking a narrow radial cut at the azimuthal angle 
   $\phi=90^\circ$. The positive relative perturbation in the stellar density at $t=1.6$~Gyr
   is shown by the dashed line.}
   \label{fig12}
\end{figure}

   A visual inspection of Fig.~\ref{fig10} and Fig.~\ref{fig11} indicates 
that the deviations from the epicycle approximation are largest near 
the convex edges of spiral arms. To better illustrate this effect, 
we take a radial cut at $\phi=90^{\circ}$ in Fig.~\ref{fig10} and Fig.~\ref{fig11}
(shown by the dashed line) and plot the resulting radial distribution of
$\xi(r,\phi=90^\circ)$ by the solid lines in Fig.~\ref{fig12}a and Fig.~\ref{fig12}b,
respectively. The positive relative stellar density perturbation $\zeta(r,\phi=90^\circ)$ along the radial
cut is shown by the dashed lines. The area below these lines is filled
with grey to identify the position of spiral arms. 
It is evident that if the local rotation velocity of stars $u_{\phi}$ is
used as a proxy for the circular speed $u_{\rm c}$ (Figure~\ref{fig12}a), then the relative errors
can become as large as $\xi=+3.0$ and $\xi=-1.0$ near the outer (convex) edges
of spiral arms. If the azimuthally averaged rotational velocity of stars $\bar{u}_\phi$
is used as a proxy for $u_{\rm c}$ (Figure~\ref{fig12}b), the corresponding relative errors 
become roughly a factor of three smaller. Yet the relative error can become
as large as $\xi=+0.7$ near the outer edge of the innermost spiral at $r\approx
15$~kpc, which renders the epicycle approximation inapplicable in these regions.

\begin{figure}
   \resizebox{\hsize}{!}{
     \includegraphics[angle=0]{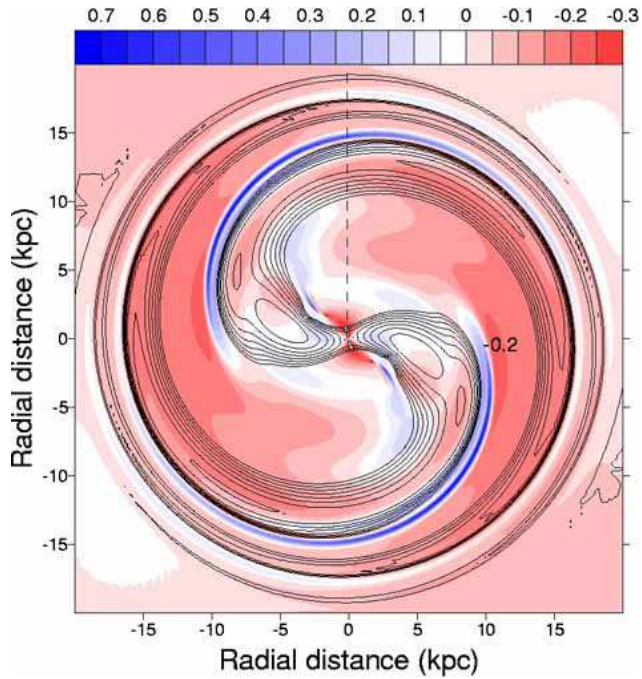}}     
   \caption{Same as Fig.~\ref{fig10} only with the circular speed determined from 
   equation~(\ref{equilib}) with the velocity dispersions set to zero.}
   \label{fig13}
\end{figure}

Finally, we check the validity of the epicycle approximation when the circular
speed is determined from the axisymmetric gravitational potential using equation~(\ref{equilib})
with $\sigma_{rr}$ and $\sigma_{\phi\phi}$ set to zero (cold stellar disc). This approach
is accurate but difficult to realize observationally for the reasons explained above.
Nevertheless, it might be instructive to compare the ratios of tangential versus
radial velocity dispersions obtained in our numerical model with those obtained from the 
exact epicycle approximation. The corresponding relative 
errors $\xi(r,\phi)$ are shown in Fig.~\ref{fig13} with the shadows of blue (positive errors) 
and red (negative errors). The spiral pattern in the disc is outlined by the
contour lines showing 
the positive relative stellar density perturbations at $t=1.6$~Gyr.
The convex edges of spiral arms are again characterized by large relative errors. 
This effect is illustrated in Fig.~\ref{fig12}c, which shows
$\xi(r,\phi=90^\circ)$ (solid line) and $\zeta(r,\phi=90^\circ)$ (dashed line) calculated along 
the radial cut at $\phi=90^\circ$ (Fig.~\ref{fig13}, dashed line). The
relative errors can become as large as $\xi=+0.7$ near the convex edge of the innermost spiral
at $r\approx 15$~kpc and are below $\pm 0.2$ elsewhere.

We conclude that the epicycle approximation is severely in error near the convex edges of spiral arms. 
On the other hand, the deviation from the epicycle approximation in the inner
disc ($r<12$~kpc) remains reasonably
small, within $10-20$ per cent. We argue that one should use the azimuthally
averaged mean tangential velocity of stars $\bar{u}_{\phi}$ (if available) 
rather than the local mean tangential velocity $u_\phi$ as a proxy for $u_{\rm c}$ 
and use the epicycle approximation with an
extreme caution near the spiral arms. It is interesting to note that the epicycle
approximation holds, to within a 20 per cent error, in and near the bar
(except for two small spots on both sides of the bar where the relative
error can become as large as $\xi=-0.3$).

\subsection{Origin of vertex deviation}

It is generally thought that two factors contribute to the vertex deviation: (i) 
moving groups and (ii) a large-scale non-axisymmetric 
component of the galactic gravitational field (Binney \& Merrifield \cite{BM}).
However, it is not clear which of the two factors is more important. The situation may become even
more complicated, since the moving groups may partly be caused by the 
non-axisymmetry in the galactic gravitational field. 
In order to get a better understanding of the origin of the vertex
deviation in our simulation, we analyse the correlation of the
vertex deviation with different dynamical quantities.


\subsubsection{Correlation of vertex deviation with the gravitational field}

\begin{figure}
   \resizebox{\hsize}{!}{
     \includegraphics[angle=0]{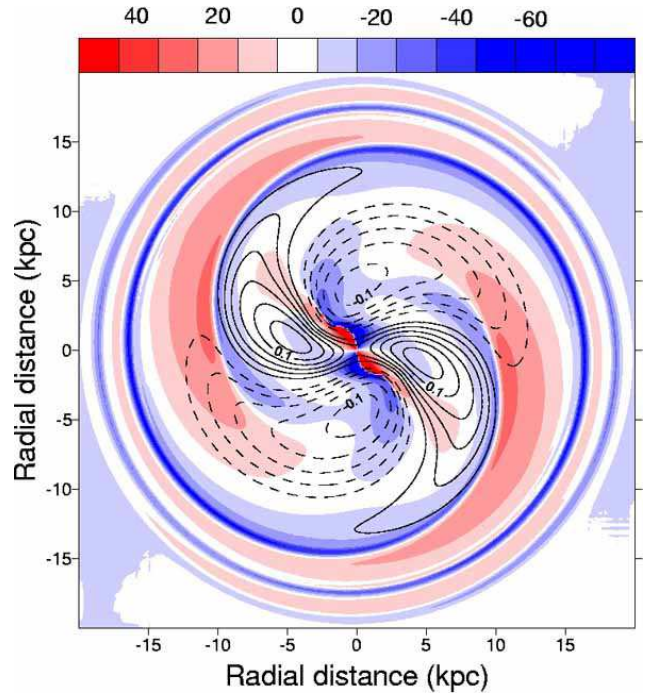}}     
   \caption{Contour lines showing the relative perturbations in the gravitational potential of the disc 
   superimposed on the spatial distribution of vertex deviations. The positive and negative perturbations
   are plotted with the solid and dashed lines, respectively. The positive and negative 
   vertex deviations are shown with the shadows of red and blue, respectively. The scale bar is in degrees.}
   \label{fig14}
\end{figure}

\begin{figure}
   \resizebox{\hsize}{!}{
     \includegraphics[angle=0]{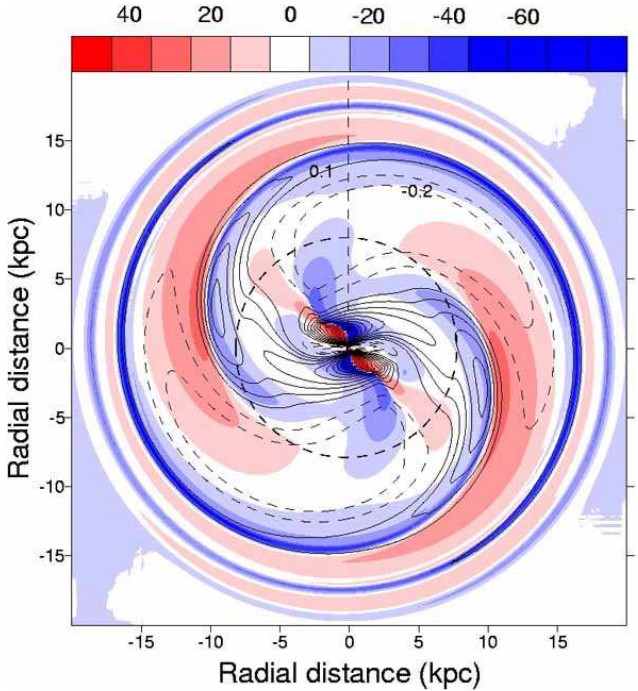}}     
   \caption{Contour lines showing the relative perturbations in the absolute value of the gravitational 
   potential gradient of the disc superimposed on the spatial distribution of vertex deviations. The positive and negative perturbations
   are plotted with the solid and dashed lines, respectively. The positive and negative 
   vertex deviations are shown with the shadows of red and blue, respectively. The scale bar is in degrees.}
   \label{fig15}
\end{figure}

We start by examining the role of the non-axisymmetric gravitational field
in generating a vertex deviation. Kuijken \& Tremaine
(\cite{KT}) considered non-axisymmetric gravitational potentials typical for a
triaxial (ellipsoidal) dark matter halo. They found that the magnitude of the vertex deviation 
should be proportional to the product of $\sin 2 \phi_{\rm b}$ and the  ratio of the non-axisymmetric 
versus axisymmetric contributions to the total gravitational potential $\epsilon_\Phi$. 
Here $\phi_{\rm b}$ is the position angle with respect to the minor axis of
the equipotential surfaces of an ellipsoidal halo (see equations [14] and [15] in Kuijken \& Tremaine \cite{KT}).
The vertex deviation is expected to be zero near the minor axis of an ellipsoidal halo 
$(\phi_{\rm b}=0)$. Of course, a direct comparison of our numerical simulations
with the analytical predictions of Kuijken \& Tremaine is problematic, 
since the dark matter halo is axisymmetric in our numerical simulations and 
the non-axisymmetry is introduced by the non-stationary stellar spiral arms 
and the bar. Moreover, the simulations allow for a completely non-linear evolution,
whereas Kuijken \& Tremaine's analysis is based on linear perturbation theory.
Nevertheless, it may be helpful to compare two different non-axisymmetric
structures and their effect on the spatial distribution of vertex deviations in a stellar
disc. 

In our model, we calculate the non-axisymmetric gravitational
potential by subtracting the gravitational potential 
of the initial axisymmetric stellar disc $\Phi^0_{\rm disc}$ from the current model's 
known gravitational potential $\Phi_{\rm disc}$ (which is the sum of contributions 
from the axisymmetric and non-axisymmetric parts of the stellar disc). 
The  ratio of the non-axisymmetric versus axisymmetric contributions to the total 
gravitational potential in our model is then determined as
\begin{equation}
\epsilon_\Phi={\Phi_{\rm disc} - \Phi^0_{\rm disc} \over \Phi^0_{\rm disc} }
\end{equation}
This quantity calculated at $t=1.6$~Gyr is shown in Fig.~\ref{fig14} by the 
contour lines. The minimum and maximum contour levels are $\epsilon_\Phi=-0.10$ and 
$\epsilon_\phi=+0.12$, respectively, and the step is 0.02.
We use dashed and solid lines to differentiate between the negative and positive 
values of $\epsilon_\Phi$, respectively. The positive and negative vertex deviations are
shown with the shadows of red and blue, respectively. It is obvious that the
magnitude of the vertex deviations in our model stellar disc {\it does not correlate } spatially 
with the ratio of the non-axisymmetric versus axisymmetric stellar gravitational potentials.
The maximal vertex deviations (by absolute value) in our model are found along the minor 
axis of the central bar, whereas in the case of an elliptical halo 
Kuijken \& Tremaine (\cite{KT}) predicted near-zero vertex deviations there.
The reason for this obvious disagreement is not clear and numerical simulations of 
vertex deviations in models with static elliptical dark matter
halos will be presented in a future paper. 

\begin{figure}
   \resizebox{\hsize}{!}{
     \includegraphics[angle=0]{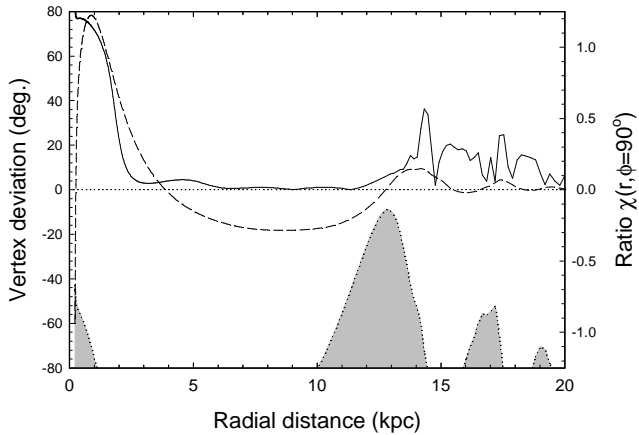}}     
   \caption{Radial profiles of the vertex deviation by absolute value $|\tilde{l}_{\rm v}|$ (solid
   line) and relative perturbation in the gravitational potential gradient
   $\chi$ (dashed line) obtained from Fig.~\ref{fig15} by taking a radial cut along
   the azimuthal angle $\phi=90^\circ$. The dotted line outlines the position
   of the bar and spiral arms by showing the positive stellar density perturbation.
   The area below the dotted line is filled with grey.}
   \label{fig16}
\end{figure}

\begin{figure}
   \resizebox{\hsize}{!}{
     \includegraphics[angle=0]{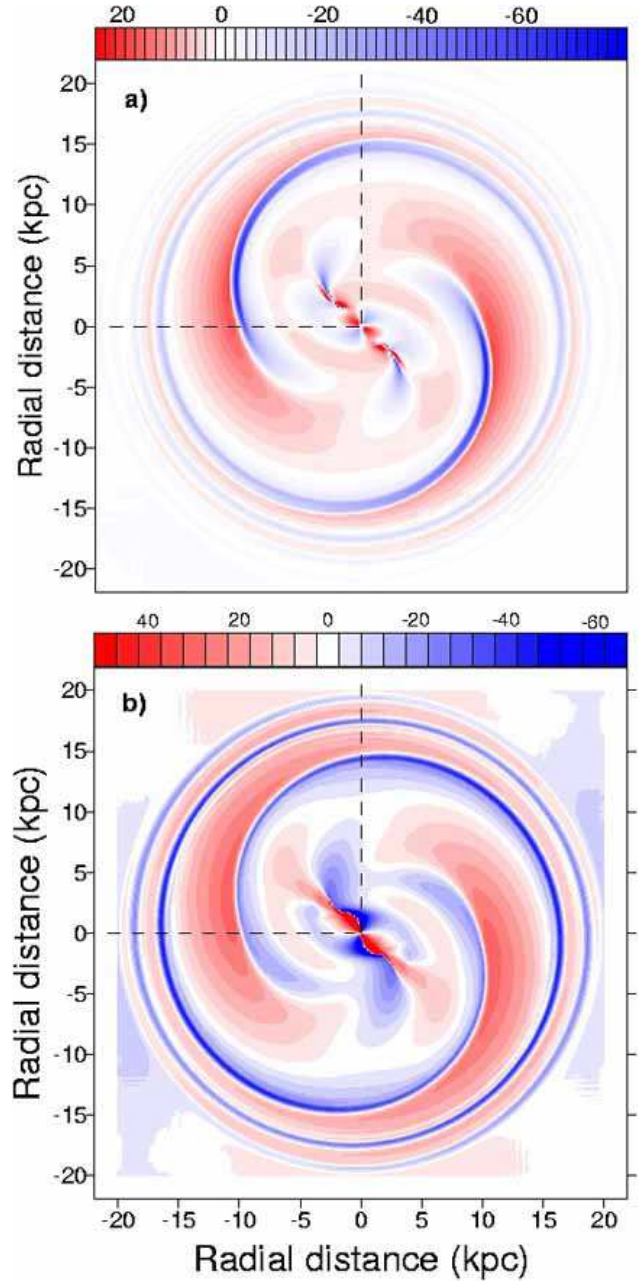}}     
   \caption{Spatial distribution of ({\bf a}) derivative $\partial u_r / \partial r$
   and ({\bf b}) vertex deviation $\tilde{l}_{\rm v}$ in the stellar disc at t=1.6~Gyr. A clear similarity is seen
   between the two quantities, especially in the outer parts of the disc. Dashed lines show radial
   cuts at $\phi=90^\circ$ (vertical) and $\phi=180^\circ$ (horizontal).
   Scale bars are in ({\bf a}) km~s$^{-1}$~kpc$^{-1}$ and {(\bf b)} in degrees. }
   \label{fig17}
\end{figure}

However, we should keep in mind that the dynamics of a star is controlled 
by the gradient of the gravitational potential rather than by the potential itself. We 
calculate the absolute value of the stellar gravitational potential gradient
in the disc as
\begin{equation}
|{\bf \nabla}\Phi_{\rm disc}| = \left[{(\partial \Phi_{\rm disc} /\partial r)^2 + (r^{-1}\partial \Phi_{\rm disc} /\partial \phi)^2 }\right]^{0.5}.
\end{equation}
This quantity can be regarded as the ``strength'' of the stellar gravitational field in the
disc. We further denote the absolute value of the initial stellar gravitational potential gradient 
$|{\bf \nabla}\Phi_{\rm disc}|_0$
and calculate the relative perturbation in the stellar gravitational potential gradient as 
\begin{equation}
\chi(r,\phi)={|{\bf \nabla}\Phi_{\rm disc}| - |{\bf \nabla}\Phi_{\rm disc}|_0
\over |{\bf \nabla}\Phi_{\rm disc}|_0 }.
\end{equation} 

The spatial distribution of $\chi(r,\phi)$ at $t=1.6$~Gyr is shown in
Figure~\ref{fig1} with the contour lines. In particular, the solid
lines delineate positive relative perturbations in the stellar gravitational potential gradient 
and dashed lines delineate negative relative perturbations. The minimum and maximum contour 
levels are $\chi=-0.3$ and $\chi=+1.2$, respectively.
The positive and negative vertex deviations $\tilde{l}_{\rm v}$ are shown with the shadows 
of red and blue, respectively. A visual inspection of Fig.~\ref{fig15} reveals a
strong correlation between $l_{\rm v}$ and positive $\chi(r,\phi)$ in the inner disc and a mild 
correlation between these quantities in the outer disc. 

To better illustrate this correlation, 
we find $\chi$ along a radial cut at $\phi=90^\circ$ 
(dashed line in Fig.~\ref{fig15}) and plot the resulting values in Fig.~\ref{fig16} 
with the dashed line. The corresponding vertex deviation is plotted with
the solid line. We should keep in mind that $\tilde{l}_{\rm v}$ changes
its sign when the galactic rotation is reversed but $\chi$ does not. Hence, we show
the absolute (rather than actual) value of the vertex deviation. The radial 
position of the spiral arms and the bar at $\phi=90^\circ$ is outlined
by the dotted line which represents 
the positive relative perturbation in the stellar surface density (the
area below this line is filled with grey). The strongest correlation between $\chi$ and 
$|\tilde{l}_{\rm v}|$ is seen in the central region where both quantities attain 
their maximum values. A noticeable correlation between $|\tilde{l}_{\rm v}|$ and $\chi$ 
is also seen near the outer (convex) edges of spiral arms at $r\approx 14$~kpc 
and $r\approx 17.5$~kpc. On the other hand, there is little or no correlation in the 
inter-arm region at $r=(15-16)$~kpc, where $\chi$ is negligible but $|\tilde{l}_{\rm
v}|$ can become as large as $20^\circ$. A similar lack of correlation is seen
at $r=(5-10)$~kpc.
We conclude that the relative perturbation in the gravitational potential gradient 
can only partly account for the magnitude of the vertex deviation in our model disc. 


\subsubsection{Correlation of vertex deviation with the streaming motions}

In this section, we search for the correlation between the vertex deviation and local 
properties of the stellar velocity field, in particular, the non-circular 
streaming motions of stars clearly seen in Fig.~\ref{fig2}. 
The streaming motion of stars can be quantified by calculating 
the spatial derivatives of the radial and tangential mean stellar  velocities, 
namely, $\partial u_r / \partial r$,
$\partial u_\phi / \partial \phi$, and $\partial u_r / \partial \phi$. 
We have examined all three characteristics of the streaming motion and found that 
$\partial u_r / \partial r$ is best suited for our purposes. 
Figure~\ref{fig17}a shows the spatial distribution of $\partial u_r / \partial r$
(in units of km~s$^{-1}$~kpc$^{-1}$) in our model stellar disc 
at $t=1.6$~Gyr. The positive and negative values of this quantity are plotted with the shadows 
of red and blue, respectively. The spatial distribution of the vertex deviation at $t=1.6$~Gyr, 
initially shown in Fig.~\ref{fig9}, is reproduced in Fig.~\ref{fig17}b to facilitate 
the comparison between $\partial u_r / \partial r$ and $\tilde{l}_{\rm v}$. 
A striking similarity between the spatial distributions 
of $\partial u_r / \partial r$ and $\tilde{l}_{\rm v}$ is clearly seen in the figure, especially 
in the outer parts of the stellar disc. 

\begin{figure}
   \resizebox{\hsize}{!}{
     \includegraphics[angle=0]{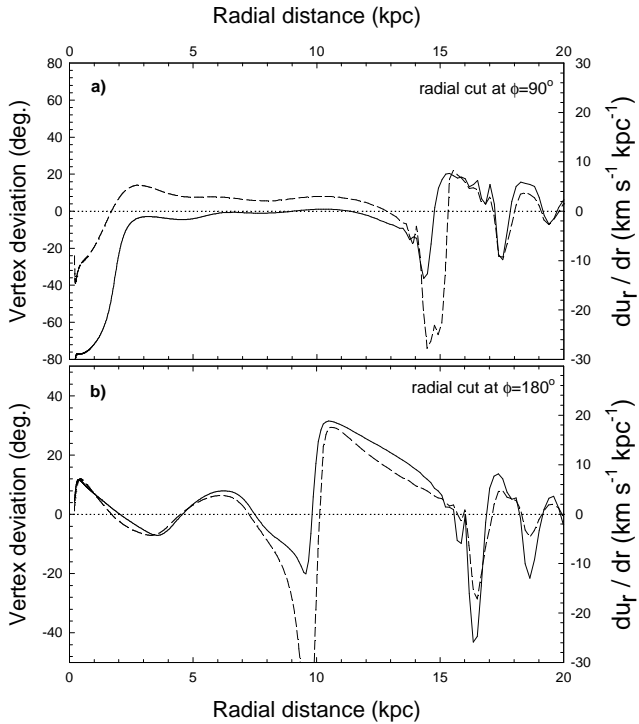}}     
   \caption{Radial profiles of the vertex deviation (solid lines) and derivative
   $\partial u_r / \partial r$ (dashed lines) obtained from Fig.~\ref{fig17}
   by taking radial cuts along the azimuthal
   angles ({\bf a}) $\phi=90^\circ$ and ({\bf b})  $\phi=180^\circ$ at
   $t=1.6$~Gyr.}
   \label{fig18}
\end{figure}

To better illustrate the spatial correlation between 
$\partial u_r / \partial r$ and $\tilde{l}_{\rm v}$, we take two radial cuts at $\phi=90^\circ$ and 
$\phi=180^\circ$ (shown with dashed lines in Fig.~\ref{fig17}) and plot the resulting 
radial profiles of $\partial u_r / \partial r$ (dashed lines) and 
$\tilde{l}_{\rm v}$ (solid lines) in Fig.~\ref{fig18}. 
Now both quantities happen to have 
the same sign and the actual values of the vertex deviation are shown.
The comparison of Fig.~\ref{fig16} and Fig.~\ref{fig18} demonstrates that 
the correlation between $\partial u_r / \partial r$ and $\tilde{l}_{\rm v}$ is in general 
better than the correlation between $\chi$ and $\tilde{l}_{\rm v}$. Yet there are
regions where $\partial u_r / \partial r$ and $\tilde{l}_{\rm v}$
show no correlation.  For instance, the derivative $\partial u_r / \partial r$ is
maximal (by negative value) at approximately 14.5~kpc (Fig.~\ref{fig18}a) but the corresponding vertex deviation 
changes its sign and is characterized by negligibly small
values there. Other characteristics of the streaming motion ($\partial u_\phi / \partial \phi$ 
and $\partial u_r / \partial \phi$) show similar correlation with the vertex deviation,
though less pronounced than in the above case with $\partial u_r / \partial r$.

To summarize, the vertex deviation is better correlated with the characteristics
of stellar streams such as $\partial u_r / \partial r$ than with
the relative perturbation in the gravitational potential gradient in the stellar 
disc\footnote{We stress that the vertex 
deviation changes its sign when the rotation is reversed but
the spatial derivatives of the mean stellar velocities do not. Hence, for a clockwise rotation,
we would have obtained an anti-correlation between $\partial u_r / \partial r$ 
and $\tilde{l}_{\rm v}$.}. 
The relative perturbation in the gravitational potential shows essentially {\it
no} correlation with the vertex deviation, which could be due to the fact
that the dynamics of stars is controlled by the gradient of the gravitational
potential rather than the gravitational potential itself. 
It appears that the non-axisymmetric gravitational field of the spiral arms and the bar perturbs stellar 
orbits by creating large-scale stellar streams. These streams introduce (under specific 
yet poorly understood conditions) the mixed velocity dispersion $\sigma^2_{r\phi}$ into 
the stellar velocity dispersion tensor and consequently generate the vertex deviation.

This result is in nice qualitative agreement with the linear perturbation
analysis in KT91. For instance, their equation (19) shows also a strong correlation between
the vertex deviation and the mean velocity field (and its gradients). Moreover, the inner
bar region is characterized by rigid rotation, which means KT91's $\alpha$ becomes unity and
the denominator formally vanishes yielding a divergent vertex deviation. Such large vertex
deviations are in good agreement with the highest $l_{\rm v}$ values 
in our simulations found near the central bar.

\subsubsection{Higher-order moments}

   When applying the Boltzmann moment equations each equation of
$n$-th order includes a term of $n+1$-order, by this formally creating
an infinite set of equations. In order to end up with a finite set
of equations, one has to introduce a closure condition. In hydrodynamics
this can be a proper equation of state (e.g.\ a polytropic equation of
state). In stellar hydrodynamics the situation is more complex, because
an obvious equation of state does not exist. For instance, an isotropic pressure
(or velocity distribution) is guaranteed in a ''normal'' gas by 
collisional relaxation which is usually a rather fast process compared to
other timescales like the dynamical timescale. Contrary to a gaseous system,
two-body relaxation in a three-dimensional stellar system is a rather slow process.
Adopting Spitzer \& Hart's (\cite{spitzer71}) formula for the
relaxation time $\tau_{\rm rel}$,
\begin{equation}
   \label{eqrelax}
   \tau_{\rm rel} = \frac{\sigma^3}{15.4 G^2 \rho \, m \ln (0.4 N)},
\end{equation}
we get with a velocity dispersion $\sigma$ of 10 km\,s$^{-1}$, a typical
stellar mass $m$ of 1 $\msun$, a local mass density 
$\rho = 0.2~\msun \, \mathrm{pc}^{-3}$ and a number $N$ of stars\footnote{Equation~(\ref{eqrelax}) 
is derived for a spherical system. Thus we
consider a sphere with a radius equal to a typical disc scale height of 100 pc
containing about $10^6$ stars. Though this is only a very rough estimate, it
does not alter our argument, because $N$ only shows up in the Coulomb logarithm
term.} of $10^6$, a relaxation time of about
$\tau_{\rm rel} \sim 1.3 \cdot 10^{12} \, \mathrm{yr}$.
This timescale exceeds the dynamical timescale (and the Hubble time) by far.
Even if we adopt the smallest values of about 7 km\,s$^{-1}$ for the
velocity dispersions of young stars, the relaxation time is still very long. 
Thus, in a purely stellar galactic disc two-body relaxation is negligible,
and heat conduction terms which are related to the third
order moments can be neglected.

  In order to check the third order moments quantitatively we started to perform
numerical test particle simulations in a given time-dependent disc potential
including a spiral. The numerical treatment is similar to the one adopted
in Blitz \& Spergel (\cite{blitz91}) and more recently in Minchev \& Quillen
(\cite{minchev07a}). First results of these simulations show that the
3rd order terms are small compared to the other moments, by this confirming
our numerical treatment. A more detailed analysis of these simulations 
will be given elsewhere.

\section{Summary and Conclusions}
\label{conclude}
   In this paper, we present two-dimensional stellar hydrodynamics simulations
of a flat galactic disc embedded in a standard dark matter halo, 
yielding a rigid rotation curve in the central part and a flat rotation curve 
outside 3~kpc.   We solve the Boltzmann moment equations up to second order
in the thin-disc approximation and follow the growth of a two-armed spiral structure  
deeply into the non-linear regime.
We analyse in detail the shape and orientation (vertex deviation) of the stellar 
velocity ellipsoids when the spiral structure has reached a saturation
phase. The simulations yield the following main results.

(i)      The shape of 
         the stellar velocity ellipsoids (as defined by the ratio $\sigma_1:\sigma_2$
         of the smallest versus largest axes of the stellar velocity ellipsoid) and 
         vertex deviations show large variations over the entire stellar disc.
         In particular, the radial distribution of the axis ratio $\sigma_1:\sigma_2$
         shows deep minima near the outer (convex) edges of stellar spiral arms.

(ii)     The spatial distributions of the radial $\sigma_{rr}$ and tangential $\sigma_{\phi\phi}$ 
         stellar velocity dispersions 
         show a similar two-armed symmetry as the 
         underlying distribution of the stellar surface density.

(iii)    The epicycle approximation fails in the regions affected by stellar
         spiral arms. In particular, it cannot reproduce the tangential-to-radial
         stellar velocity dispersion ratio $\sigma_{\phi\phi}:\sigma_{rr}$
         obtained from the numerical simulations. The deviations from the epicycle approximation 
         are especially large near the outer (convex) edges of stellar spiral arms, 
         where the epicycle approximation
         may be in error by a factor of two or more.
 
(iv)    Globally, there is an excellent correlation between the mass-weighted
         vertex deviation (integrated over the whole stellar disc) and the
         global Fourier amplitude of the stellar density perturbations.
         However, locally, there is no simple correlation between
         the vertex deviation and the mass distribution.
 
(v)     The magnitude of the vertex deviation is not correlated
         with the gravitational potential (mass distribution) and is only weakly correlated
         with the gradient of the gravitational potential.
         However, the vertex deviation correlates strongly with the spatial
         gradients of mean stellar velocities, in particular, with the
         radial gradient of the mean radial velocity, $\partial u_r/\partial
         r$
         
(vi)    Large-scale, non-circular stellar streams induced by the non-axisymmetric gravitational
        field of the stellar spiral arms and the bar may be responsible
        for the failure of the epicycle approximation, large vertex deviations,
        and peculiar shapes of the stellar velocity ellipsoids.

 Our results are in good qualitative agreement with the results of the
linear perturbation analysis given in KT91. Moreover, our simulations include also
the non-linear stages of the disc evolution when dynamical feedback limits the growth of
a spiral structure. The large spatial variations of the properties of the stellar
velocity distribution found in our numerical simulations and the mismatch to the epicycle approximation
are a caveat for any analysis of observational data which cover larger
spatial areas or which include stellar samples coming from a larger spatial region.
 It will be interesting to check whether features like the ''u-anomaly'' could also be
created by large-scale streams induced by a transient or swing-amplified
spiral.

\section*{Acknowledgements}
We thank the referee, Ivan Minchev, for useful suggestions that helped
improve the manuscript.
This project was partly supported by grants RFBR 06-02-16819-a, South Federal
University grant 05/6-30, and Federal Agency of Education 
(project code RNP 2.1.1.3483). E.I.V. acknowledges support from a CITA National Fellowship.
Numerical simulations were done on the Shared Hierarchical Academic Research Computing Network (SHARCNET).
The authors are grateful to Panos Patsis for stimulating discussions.

\end{document}